\documentclass[draft]{agujournal2018}
\usepackage{apacite}
\usepackage{url} 
\newcommand{\fig}[1]{Figure~\ref{fig_#1}}
\newcommand{\Bsw}{B_{\mathrm{SW}}}
\newcommand{\Bs}{B_{\mathrm{sheath}}}
\newcommand{\Bm}{B_{\mathrm{ME}}}
\newcommand{\Bw}{B_{\mathrm{wake}}}

\newcommand{\Vs}{V_{\mathrm{sheath}}}
\newcommand{\Vm}{V_{\mathrm{ME}}}
\newcommand{\Vw}{V_{\mathrm{SW}}}

 \newcommand{\sect}[1]{Section~\ref{sect_#1}}

\newcommand{\adv}{    {Adv. Spa. Res.}}
\newcommand{\aap}{Astron. Astrophys.}
\newcommand{\apj}{Astrophys. J.}
\newcommand{\grl}{    Geophys. Res. Lett.}
\newcommand{\jgr}{J. Geophys. Res.}
\newcommand{\nat}{Nature}
\newcommand{\ssr}{Space Sci. Rev.}
\newcommand{\solphys}{{\it Solar Phys.}}

\newcommand{\annG}{   {Annales Geophysicae}}
\newcommand{\jastp}{  {J. Atmos. Sol. Terr. Phys.}}
\newcommand{\planss}{ {\it Planetary Spa. Sci.}}

\draftfalse

\journalname{JGR: Space Physics}

\begin{document}

%
%


\title{Generic {magnetic field intensity} profiles of interplanetary coronal mass ejections at Mercury, Venus and Earth from superposed epoch analyses}

%
%




\authors{Miho Janvier\affil{1}, Reka Winslow\affil{2}, Simon Good\affil{3}, Elise Bonhomme\affil{1}, Pascal D\'emoulin\affil{4}, Sergio Dasso\affil{5,6}, Christian M\"ostl\affil{7}, No\'e Lugaz\affil{2}, Tanja Amerstorfer\affil{7}, Elie Soubri\'e\affil{1}, Peter D. Boakes\affil{7}}


\affiliation{1}{Institut d'Astrophysique Spatiale, CNRS, Univ. Paris-Sud, Universit\'e Paris-Saclay, B\^at. 121, 91405 Orsay Cedex, France.}

\affiliation{2}{Institute for the Study of Earth, Ocean, and Space, University of New Hampshire, Durham, NH, USA}

\affiliation{3}{Department of Physics, University of Helsinki, Helsinki, Finland}

\affiliation{4}{LESIA, Observatoire de Paris, Universit\'e PSL, CNRS, Sorbonne Universit\'e, Univ. Paris Diderot, Sorbonne Paris Cit\'e, 5 place Jules Janssen, 92195 Meudon, France}

\affiliation{5}{Instituto de Astronom\'\i a y F\'\i sica del Espacio, UBA-CONICET, CC. 67, Suc. 28, 1428 Buenos Aires, Argentina}

\affiliation{6}{Departamento de Ciencias de la Atm\'osfera y los Oc\'eanos and Departamento de F\'\i sica, Facultad de Ciencias Exactas y Naturales, Universidad de Buenos Aires, 1428 Buenos Aires, Argentina}

\affiliation{7}{Space Research Institute, Austrian Academy of Sciences, A-8042 Graz, Austria}





\correspondingauthor{Miho Janvier}{miho.janvier@ias.u-psud.fr}




\begin{keypoints}
\item Slow ICMEs have a more symmetric profile compared with fast ICMEs. This trend is maintained at different heliospheric distances.
\item ICMEs sampled at Mercury have smaller sheaths and magnetic ejecta than further away. As they propagate, their sheath sizes increase.
\item At all three spacecraft, the post-ICME solar wind does not fully recover its original properties, indicating a long recovery period.
\end{keypoints}

%
%


\begin{abstract}
We study interplanetary coronal mass ejections (ICMEs) measured by probes at different heliocentric distances (0.3-1AU) to investigate the propagation of ICMEs in the inner heliosphere and determine how the generic features of ICMEs change with heliospheric distance.
Using data from the MESSENGER, Venus Express and ACE spacecraft, we analyze with the superposed epoch technique the profiles of ICME substructures, namely the sheath and the magnetic ejecta.
We determine that the median magnetic field magnitude in the sheath correlates well with ICME speeds at 1 AU and we use this proxy to order the ICMEs at all spacecraft. We then investigate the typical ICME profiles for three categories equivalent to slow, intermediate and fast ICMEs.
Contrary to fast ICMEs, slow ICMEs have a weaker solar wind field at the front and a more symmetric magnetic field profile. We find the asymmetry to be less pronounced at Earth than at Mercury, indicating a relaxation taking place as ICMEs propagate.
We also find that the magnetic field intensities in the wake region of the ICMEs do not go back to the pre-ICME solar wind intensities, suggesting that the effects of ICMEs on the ambient solar wind last longer than the duration of the transient event. 
Such results provide an indication of physical processes that need to be reproduced by numerical simulations of ICME propagation. The samples studied here will be greatly improved by future missions dedicated to the exploration of the inner heliosphere, such as Parker Solar Probe and Solar Orbiter.
\end{abstract}

%
%

%
%

\section{Introduction}
\label{sect_Introduction}
  The Sun is the source of intermittent ejections of bulk plasma and magnetic field structures called coronal mass ejections \citep[CMEs, see review in][]{Webb2012}. With space probes, these structures are observed as they propagate in the interplanetary medium. Interplanetary coronal mass ejections, or ICMEs  \citep{Kilpua2017} that have been detected by in situ instruments, are identified by a combination of plasma and magnetic field parameters {{\citep{Klein1982,Jian2006,Wimmer2006}}}.  
  
   {{One or a combination of plasma characteristics different than the ambient solar wind (SW) can be observed in ICMEs. When ICME fronts propagate faster than the ambient SW, a shock forms, and the accumulated SW material between the shock and the ejecta is called the ICME sheath. In the sheath, the magnitudes of the density and the magnetic field increase, while the magnetic field variance is large. Behind the ICME sheath is a magnetically dominated region with less intense magnetic fluctuations than in the sheath, which we term magnetic ejecta (ME), similarly to {\citet{Winslow2015, winslow2016}}. The subset of MEs that exhibit rotation in the azimuthal magnetic field component, and have proton temperatures and plasma $\beta$ (ratio of the plasma thermal pressure to the magnetic pressure) values that are lower than the ambient SW, are termed magnetic clouds \citep[hereafter MCs,][]{Burlaga1981}. {These MCs are typically modelled with twisted magnetic field configurations, or flux ropes (\citealp{Dasso09b} \citealp{AlHaddad2013})}. Given that plasma parameters are not available from the MESSENGER and Venus Express missions, identifying MCs is not possible in these data, and thus we adopt the generic term ME to refer to the ICME substructure that is magnetically dominated. In this paper, we refer to an ``ICME" as a set of two substructures: ``sheath" and ``ME", whereas we note that {another common} convention termed these substructures ``sheath" and ``ICME".}}

ICMEs propagate within the heliosphere and their arrival and consequences at the planets of the solar system can be tracked {{\citep[e.g. see][]{Cane2000, Prange2004, Witasse2017}}}. ICMEs are known to generate geomagnetic storms at Earth (e.g., \citealp{Gosling1990}, \citealp{Lindsay1995}, \citealp{Farrugia1997}, \citealp{Zhang2004}), while ICME-related space weather events have been observed at Mercury \citep{Slavin2014, Winslow2017} and Mars as well \citep{Lee2017}. 

In \citet{Masias2016}, the authors investigated how the generic profiles of ICMEs {with well-defined MCs appear} at 1~AU, using
 measurements of the solar wind transients by the ACE spacecraft.
{The authors used a superposed epoch analysis to derive the generic profile of ICMEs at 1~AU.   They found that this profile changed depending on the speed of ICMEs.} By creating subgroups of ICMEs ranked with their speeds, they found that ICMEs with the slowest speeds have a symmetric generic magnetic field profile, while the subgroup with the fastest speeds have a non-symmetric profile, with a stronger magnetic field in their front than in their rear. The symmetric profiles were interpreted as evidence of a relaxation mechanism taking place: as the magnetic field pressure within the MC tries to be in balance with the surrounding solar wind, one would expect a more symmetric profile with a lower propagation speed than an MC propagating faster than the SW, therefore accumulating a sheath with stronger magnetic field and plasma pressures \citep{Liu2008}. The presence of a strong sheath then implies surrounding conditions at the front and at the wake of the ME different from one another.

Investigating ICMEs seen at different distances {can} provide important information about how these structures behave during their propagation in the heliosphere and interact with different planets' space environments (\citealp{Bothmer1998}, \citealp{Liu2005}, \citealp{Good2015}, \citealp{Winslow2015}, \citealp{Good2018}).

{A superposed epoch analysis (SEA), also known as a Chree analysis \citep{Chree1914}, provides a mean to statistically analyze patterns in a {time-varying parameter in a population of events}. {A classical SEA typically involves the use of a characteristic reference time, such as the event starting time, to align all the studied events to a common zero epoch time.  Then, events are binned to a common set of time bins and averaged in each time bin to deduce a mean or median temporal profile. When the studied events also have clearly defined end times, the event timescales can be normalized (e.g. on a scale from 0 at the event start to 1 at the event end) as part of the SEA, prior to the binning and averaging.}
By averaging the profiles, the superposed epoch analysis reinforces the common features of the events. SEA is a well established technique that is regularly used in a wide range of disciplines, including in the geophysics and space physics community; for example, it has been used to infer the generic properties of co-rotating/stream interaction region profiles in order to make comparisons with ICMEs \citep{Yermolaev2017}, and it has been used to analyze ICME substructures (\citealp{Klein1982}, \citealp{Zhang1988}, \citealp{Rodriguez2016}).}

It is important to note that the SEA technique should only be applied to {a set of events with features that appear to show some ordering in time in the parameter used in the analysis.} If applied to parameters without the same time ordering (e.g. the ICME north-south field component), significant features may be averaged out.



In the following, we investigate how the propagation away from the Sun affects the generic profile of ICMEs. For this study, we therefore consider three different samples of ICMEs observed at spacecraft positioned at different heliospheric distances. The observations made near Mercury's orbit are {provided by the MESSENGER mission and those made near Venus's orbit are provided by the Venus Express mission; we also analyze events observed by the ACE spacecraft that were previously studied by \citet{Masias2016}.}\\

We first start in \sect{datasets} by describing the different datasets used in the paper and their characteristics. {Since MESSENGER and Venus Express were planetary spacecraft that regularly crossed in and out of the magnetosphere and solar wind, identifying ICMEs using these spacecraft datasets has required more careful consideration.} In \sect{SE}, we introduce the superposed epoch analysis technique before applying it to the different samples we have at the three different space missions. The results are shown in \sect{overallSE}, where we look at the superposed epoch analysis on the whole datasets at different spacecraft, before searching for a method to create subcategories of events and the specific profiles of these subcategories in \sect{subcategories}. Discussions and conclusions are given in \sect{conclusion}.

%
%

\section{Description of the datasets}
\label{sect_datasets}

In the present study, we use the data from three planetary missions, MESSENGER, Venus Express and ACE. A brief description of the missions as well as their datasets is given below.\\

\subsection{The MESSENGER mission and data}

\subsubsection{Description of the dataset}

The MErcury Surface, Space ENvironment, GEochemistry, and Ranging (MESSENGER) mission \citep{Solomon2007} was launched on 3 August 2004 and terminated on 30 April 2015. After multiple flybys, the spacecraft reached Mercury for its orbit insertion on 18 March 2011. 

The Magnetometer \citep[MAG][]{Anderson2007} onboard MESSENGER operated continuously after 2007, during the cruise phase of the mission with a typical sample rate of 2~s$^{-1}$. It has thus produced a database of interplanetary magnetic field (IMF) observations in the inner heliosphere in the heliocentric distance range of $0.3 < r < 0.6$~AU. After its orbit insertion around Mercury, due to its highly eccentric orbit, MESSENGER spent 40\% to 85\% of its time outside of Mercury's magnetosphere, in the solar wind. 
Although the MESSENGER payload included a plasma spectrometer (the Fast Imaging Plasma Spectrometer \citep[FIPS][]{Andrews2007}, the spacecraft was three axis-stabilized and FIPS had a limited field of view that did not allow for the recovery of the solar wind density. Solar wind speed and temperature could be derived from the measurements about 50\% of the time that MESSENGER was in the solar wind \citep{Gershman2012}. 
{ICMEs have been identified in MESSENGER data near Mercury's orbit since early January 2009 \citep{Good2016}.} The orbital phase of the mission coincided with the ramp up to and the maximum of the solar cycle 24. This results in a number of ICMEs observed during MESSENGER's orbital phase (after 2011) being higher than that in during the cruise phase.

MESSENGER magnetic field data are available on the Planetary Data System (\url{https://pds.nasa.gov/}), and were made accessible for this study through the HELCATS project {(\url{https://www.helcats-fp7.eu/}; \citet{Mostl2017})}. Due to MESSENGER's crossing in and out of Mercury's magnetosphere during the orbital phase of the mission, the data needed to be cleaned, i.e. the magnetospheric passages had to be removed to leave behind only interplanetary observations. For any given ICME, we have selectively included sections of data where MESSENGER is clearly outside of the magnetosphere, specifically, outside of the bow shock boundary. The magnetospheric boundaries move faster than the spacecraft, leading to MESSENGER crossing the bow shock multiple times, both before and after crossing the magnetosphere. In order to maximize the available data for the ICME profiles, we have identified all the bow shock crossings during each ICME event and have included all data sections outside of those boundaries, even sections of a few minutes in length. For details on identification of the bow shock boundary in the MESSENGER data the reader is referred to \citet{Winslow2013}.

\subsubsection{Description of the ICME catalogs}

The list of {MESSENGER} ICMEs studied here comes from two {catalogs}. The first catalog of {events} is from \citet{Winslow2015}, later completed until the end of the MESSENGER mission \citep{Winslow2017}, with a total of 69 ICMEs. The time interval (2011-2015) corresponds to the {phase} of the mission when MESSENGER was in orbit around Mercury. The authors selected ICMEs based on several criteria: having a clear interplanetary discontinuity {seen as a step-function-like magnetic field increase} followed by a sheath, a magnetic ejecta (ME), and causing a visible distortion of Mercury's magnetosphere {\citep[see details in][]{Winslow2015}}. {This implies that the catalog lists strong MEs that drove interplanetary shocks. However, the exact nature of the discontinuities (e.g. whether they were shocks or waves) could not be determined given the lack of key plasma parameters.} Here, the term ``magnetic ejecta'' is used to identify a strong and smooth magnetic field region, while a rotation {(generally indicative of a flux rope, see \sect{Introduction}) is not always present.}

The second catalog of events is from \citet{Good2016}. This {list includes} 36 magnetic clouds and ICMEs detected by MESSENGER during the period 2005-2012, therefore covering mostly the cruise phase. Note that out of these 36 events, only 20 events were within a reasonable distance of Mercury's average orbit {of 0.395 AU (an heliospheric distance range of 0.309 AU $< r <$ 0.463 AU was selected)}, out of which 10 events were also found in the \citet{Winslow2015} catalog. Note that {the small number of events in common} is due to the fact that \citet{Good2016} {selected only} well-defined magnetic clouds, i.e. with a relatively clear and smooth rotation of the magnetic field direction, coinciding with a relatively enhanced field magnitude {compared with the solar wind magnitude level}, over a period of at least 4 or 5 hours. {These signatures are sometimes difficult to discern in data} perturbed by magnetospheric crossings, therefore limiting the number of detected cases during the nominal phase of the mission at Mercury's orbit. 


\subsubsection{Selecting events}

\begin{figure}[t!]    
\centering
\includegraphics[width=1\textwidth, clip=]{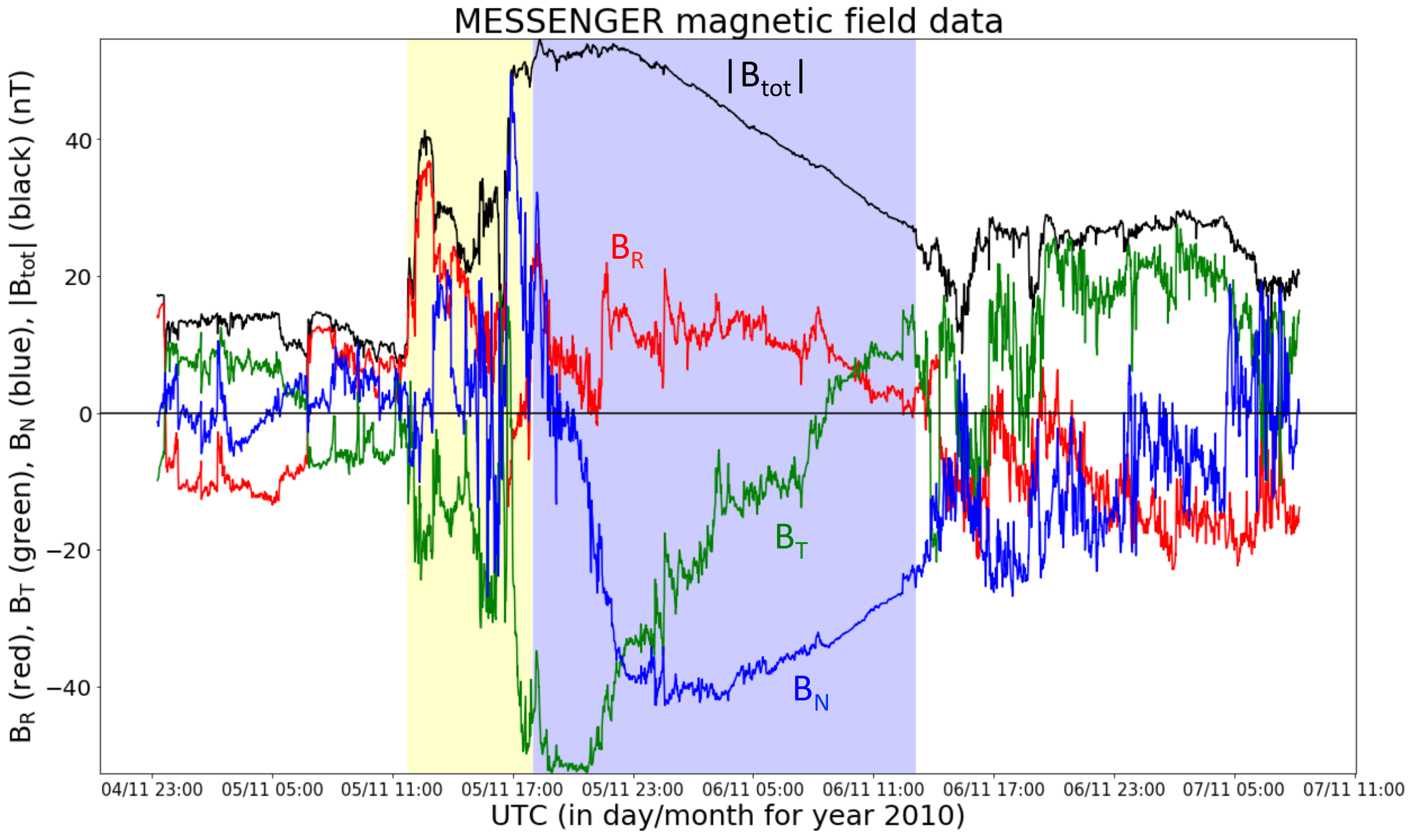}
\caption{Example of an ICME detected by MESSENGER {during the cruise phase, at distance $r$=0.462~AU}. The $x$-axis represents the time, while the $y$-axis represents the intensity of the magnetic field. The presence of a sheath is indicated by the yellow area while the magnetic ejecta is bounded by the blue area. The black line traces the total magnetic field. The $B_T$ (in green) and partly $B_N$ (in blue) components show a coherent rotation, typical of an azimuthal/poloidal flux rope component, while $B_R$ (in red) and partly $B_N$ (in blue)  show a stronger field in the central region than at the {boundaries}, again typical of {the axial field component within a} flux rope. This behavior indicates the presence of a flux rope.}
\label{fig_ExampleICMEMES}
\end{figure}

Out of these 79 ICMEs, some had the interplanetary shock or magnetic ejecta boundaries within Mercury's magnetosphere, others had data {gaps that} were too large, so that these events could not be used for the purposes of the present study. Overall, we constrained our data sample to 41 ICMEs that presented enough data within each substructure (at least $\sim 60\%$ of data coverage within the sheath/ME), had clear signatures of the sheath and ME, and had the boundaries of each substructure clearly defined. {An example of an ICME} detected at MESSENGER is shown in \fig{ExampleICMEMES}{, where we have represented a time-series of the intensity of the magnetic field (the total magnetic field in black, $B_R$ in red, $B_T$ in green and $B_N$ in blue). The shaded area in yellow indicates the presence of a sheath region, while the blue area indicates the presence of an ME.} 
The ICME shown in \fig{ExampleICMEMES} corresponds to when MESSENGER was still in cruise phase (but with the spacecraft positioned at $r=0.462$~AU, i.e. close to Mercury's aphelion). 
{Examples of ICMEs interacting with Mercury's magnetosphere are} given in \citet{Winslow2015} and \citet{Good2015}.

  \subsection{The Venus Express Mission}

	\subsubsection{Description of the dataset}
Venus Express \citep[][VEX in the following]{Titov2006} was an ESA mission launched on 9 November 2005 and officially ended in December 2014. It was inserted in orbit around Venus on 11 April 2006. Similarly to the MESSENGER mission, VEX spent enough time in the solar wind during its orbit around the planet to allow the detection of ICMEs. The VEX magnetometer \citep[MAG,][]{Zhang2006} detected its first ICME in July 2006 and continued its sampling of solar wind structures up to the end of the mission. The orbital phase of the mission {overlaps with} the time interval of MESSENGER orbiting around Mercury, and coincides with the {solar minimum between cycles 23 and 24, and with the ascending and maximum phases of cycle 24.} 

VEX's magnetometer data used in the present study are available in the ESA's Planetary Science Archive (at \url{https://www.cosmos.esa.int/web/psa/venus-express}, and were made accessible for this study through the HELCATS project. They correspond to a 1-min cadence sampling. The magnetic field was perturbed by magnetospheric crossings of the planet. Although Venus does not generate its own magnetosphere, an induced magnetic field arises from the interaction between the solar wind and the planet's ionosphere. Then, each magnetospheric crossing was cleaned in order to render a solar-wind-only dataset for ICME analysis. On average, there is one magnetospheric crossing per 24 hours {with a typical duration of 2.5 hours} throughout the VEX data, while an ME is typically crossed in 14 hours. 

	\subsubsection{Description of the ICME catalogs}
We used the VEX ICME catalog provided by \citet{Good2016} where 84 events are listed, from 2006 to 2013.  This catalog focuses on ICMEs that have a clear magnetic ejecta signature. This means that even though plasma parameters are not available from the mission, the authors identified time intervals where the magnetic field strength relative to that of the ambient solar wind field is enhanced, for at least a period of 4 hours. While this may seem relatively short compared with the typical duration of ICMEs near Earth (approximately 1 day, see Table \ref{tableduration}), this threshold was allowed to account for the expansion of ME as they move away from the Sun. The authors also {inferred the presence of a magnetic cloud whenever there was a magnetic field rotation}.  However, the lack of plasma measurements for the VEX mission, as for MESSENGER, makes it impossible to define a proper magnetic cloud: we therefore only define ME for both of these two missions.

\begin{figure}[t!]    
\centering
\includegraphics[width=1\textwidth, clip=]{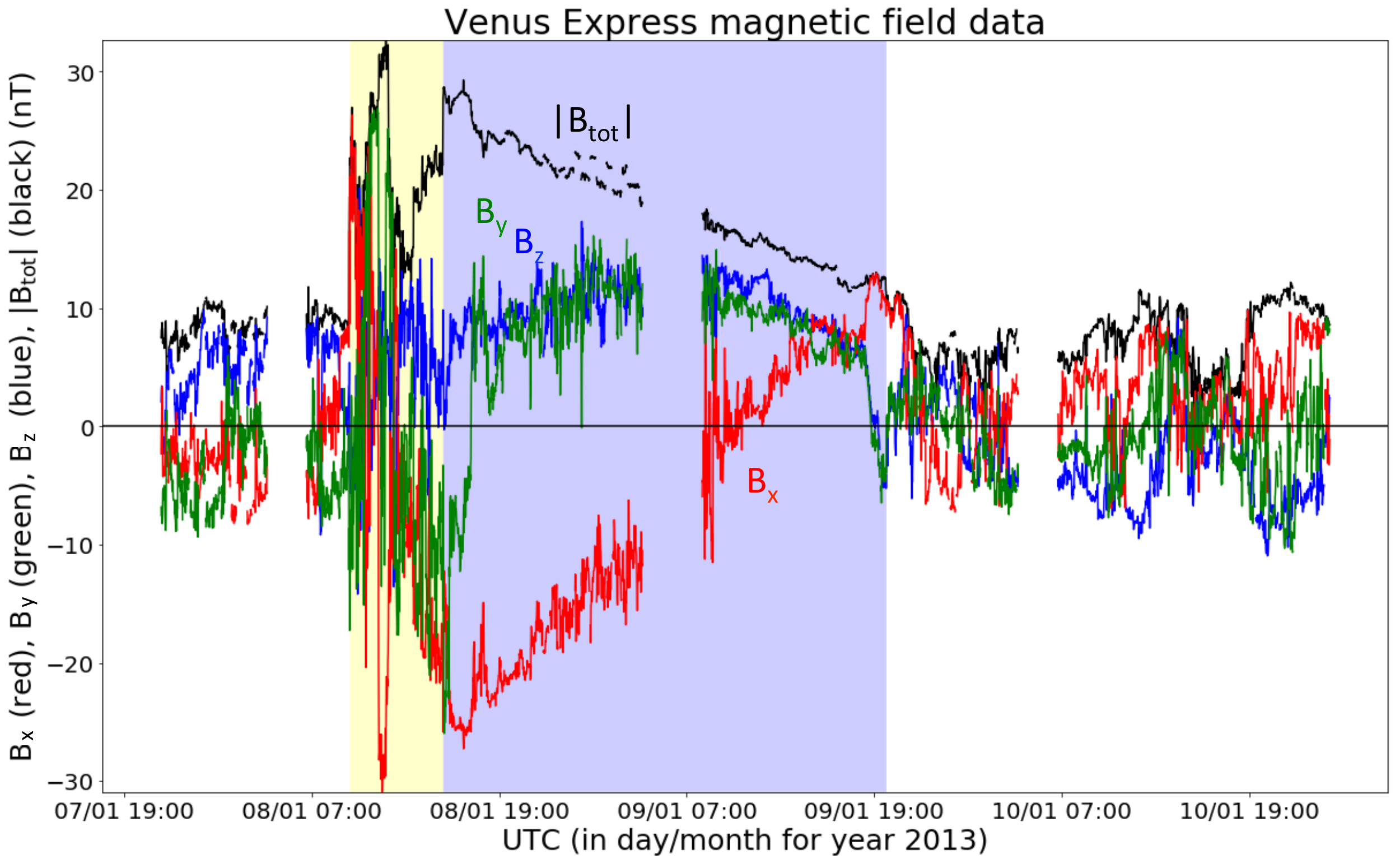}
\caption{Example of an ICME detected by Venus Express. The format of the graph ($x$ and $y$-axis) as well as the choice of colour for the lines are the same as in \fig{ExampleICMEMES}. The coordinates for the magnetic field vector are expressed in Venus Solar Orbital (VSO) co-ordinates: x points from Venus towards the Sun, y from Venus in the direction opposite to the planetÕs orbital motion, and z completes the right-handed set. Here, one of the magnetic field components ($B_x$ in red) shows an inversion of sign {in the ME (blue shaded area)}, while the two others are maximum near the center. They are the indication of the presence of a flux rope. The absence of data in the middle of the magnetic ejecta is due to the magnetospheric crossing.}
\label{fig_ExampleICMEVEX}
\end{figure}

Since the data were cleaned for induced magnetospheric passes, the detection of the ICME sub-structure boundaries is not straightforward, and only 67 of the \citet{Good2016} events were used in the analysis discussed in this paper.
An example of an ICME found in the Venus Express data is shown in \fig{ExampleICMEVEX}. We identified 19 ICMEs that did not have a sheath region.

  \subsection{ACE Mission}

The data collected at 1AU are provided by the Advanced Composition Explorer \citep[ACE][]{Stone1998}. This spacecraft was launched on 25 August 1997 and has a stable orbit around the L1 libration point. The spacecraft provides a continuous coverage of solar wind parameters. Similarly to \citet{Masias2016}, we use the 1-s time cadence interplanetary magnetic field and 64-s time cadence plasma measurements from two of the instruments aboard the spacecraft (MAG and SWEPAM experiments, see \citealp{Smith1998}, \citealp{McComas1998}). The data from ACE are directly available online at \url{http://www.srl.caltech.edu/ACE/ASC/level2/}. 

For consistency and to make comparisons possible with the study of \citet{Masias2016}, we choose to consider their list of the 42 events taken from their Table A.1 (44 are listed in their paper, however data is only available for 42 events in the ACE MAG database). These events correspond to ICMEs detected between 1998 and 2006 that have been flagged in the ICME list of \citealt{Richardson2010} as MCs, which \citet{Masias2016} crossed with the MC list from Lepping (\url{http://lepmfi.gsfc.nasa.gov/mfi/mag\_cloud\_S1.html} \citep{Lepping2006}. Here, all ICMEs are associated with the presence of a sheath and an associated shock.  Note that here, all ICMEs have a well defined MC, including the plasma characteristics (\sect{Introduction}) that can be observed because ACE solar wind plasma data are available.
While this provides a bias for the ACE list {(since most of the ICMEs at 1~AU do not have a clearly identified MC)}, the idea is to use ACE results as a guideline to get more understanding on how to analyze the superposed epochs for the MESSENGER and VEX ICMEs. 

  \subsection{List of the events with revised boundaries}
	\label{sect_revisedboundaries}
	
The identification of ICME boundaries is highly dependent on the set of parameters used (i.e. plasma vs magnetic field measurements) as well as on the authors making the identification: see, e.g. the study by \citealp{Riley04}, \citealp{Richardson2004}, \citealp{Russell2005}, \citealp{Dasso06}, \citealp{AlHaddad2013}, where the authors analyzed this dependency. Thus, we have analyzed the times of the ICME boundaries again and have provided our assessments of the boundary times for MESSENGER, VEX and ACE (see Section~\ref{boundaries_analysis}). In catalogs of ICMEs detected by {MESSENGER or VEX, the leading edge of the sheath was identified by an increase in the magnetic field intensity and the field variance.} For ACE, the addition of plasma parameters provides another constraint in the definition of the boundaries of the ICME substructures. 
However, identifying the trailing edge for all these ICMEs is complicated, as plasma parameters and magnetic field components, when available, may not show a distinct transition when the spacecraft crosses the interface between the ME and the solar wind.
Furthermore, events that are found in both \citet{Winslow2015} and \citet{Good2016} MESSENGER catalogs do not necessarily give the same boundaries for the sheath and the magnetic ejecta. Our updated catalogs are made available in Appendix~A.


%
%


\section{Superposed Epoch Analysis}
\label{sect_SE}

\subsection{Normalization and binning}
\label{sect_normbin}

We use the superposed epoch analysis in the following as a statistical method to determine averaged time profiles of physical parameters obtained in situ. Supposing that these physical parameters behave in a similar fashion, averaging time series from all events on a normalized time scale provides meaningful mean and median profiles. Doing so accentuates the typical features. The method used to prepare the data for the superposed epoch is the same as in \cite{Masias2016}. 

In order to obtain a superposed epoch for all events, the time series data for each event, is first normalized so that the sheath start time (or the discontinuity time), and the magnetic ejecta (ME) start and end times are all normalized to the same new time range. When defining the normalized time scale, one can freely choose the interval length for both the sheath or the ME. However, these two intervals are not independent from each other, as was shown in \cite{Masias2016} where the authors found that the interval size ratio between the physical length of the ME interval and the sheath was close to 3 for ACE data, meaning that the size of the magnetic ejecta was on average three times longer than the size of the sheath at 1~AU.  Thus, we use two different time ranges as normalized times, one range for the sheath and another one for the ME.


Once the time series are normalized, the time intervals are divided by the same number of bins for each event. This is because each ICME does not have the same duration nor the same number of data points, so that by binning the data, we get the same number of data points for each event normalized on the same timescale. In the present case, we impose 50 time bins for the sheath interval, while 150 bins are taken for the ME. The superposed epoch also includes {an interval of solar wind prior to the arrival of the sheath with a duration twice that of the sheath, and a wake region with a duration equal to that of the ME}. Then, the data from different events but of the same bin number are added together and the mean and median values of the magnetic field intensity in each bin are calculated. The data gaps are taken into account for each bin implying that the means and medians are computed with a variable number of cases {along the normalized time}.

\subsection{The size ratio of ICME substructures}
\label{sect_size_ratio}

The interval size ratio between the physical length of the ME interval and the sheath ratio may change depending on the ICME speed. However, it is on average similar for slow and fast ICMEs, as is shown in Fig.1 of \citealt{Masias2016}. 
Hence, since the superposed epoch aims at providing the most likely generic profile, it is important to use the typical interval size ratio when doing the normalization.

Since there are no continuous plasma parameters (and hence, continuous solar wind speed measurements) for MESSENGER and for VEX, the spatial size of the magnetic ejecta as well as the sheath cannot be inferred for events seen at these two spacecraft. Indeed, the velocity information is needed to transform the time range observed into an associated spatial size. We then estimated the size ratio of ICMEs at all spacecraft, including ACE, by comparing the time interval length for the sheath and the ME, rather than the actual physical size.   

First, we calculated the median duration for all ICMEs that had both a sheath and an ME clearly defined in all MESSENGER, VEX and ACE data. We use the median value because it is statistically more robust and resistant to outliers than the average value. The results are reported in Table \ref{tableduration} and are expressed in units of day. We find that on average, both the sheath and the ME have a smaller duration near Mercury (0.1 and 0.3 days respectively) than near Earth (0.5 and 0.8 days respectively). Assuming no large deceleration of ICMEs between Mercury's orbit and 1 AU, this confirms the numerous studies that have shown that the magnetic substructure of ICMEs expands away from the Sun \citep[e.g.][and references therein]{Klein1982, Gulisano2010, Nakwacki2011, Gulisano2012}. 

\begin{table}[h!]
\centering
\begin{tabular}{| c || c | c | c |}
\hline
\multicolumn{4}{|c|}{Duration of sheath and ME in ICMEs at different heliocentric distances}\\
\hline
 &  MESSENGER ($\approx 0.4$~AU)& Venus Express (0.72~AU)& ACE (1~AU)\\
\hline
Sheath duration & 0.1 days & 0.3 days & 0.5 days\\ 
ME duration & 0.3 days & 0.6 days & 0.8 days\\
ratio&  0.33  & 0.38 & 0.66\\
\hline
\end{tabular}
\caption{Median values for the duration (expressed in days) of the sheath (top line) and the ME (middle line), and ratio of the sheath and ME durations (bottom line) for all ICMEs seen at MESSENGER, VEX and ACE}
\label{tableduration}
\end{table}

The ME to sheath duration ratio increases from Mercury to Earth. This is because the rate of size increase (i.e., expansion) is typically different between ME and sheaths.
We find that between Mercury's and Venus' orbits, the sheath duration increases 3 times, compared to 1.7 times between Venus' and Earth's orbits. 
 {However,} the duration of the ME increases 2 times between Mercury's and Venus' orbit, and 1.3 times between Venus' and Earth's orbit. Both substructures see a bigger increase in duration between Mercury's and Venus' orbits compared with Venus' and Earth's orbits, which is expected since Venus is closer to Earth than it is to Mercury.

While both the sheath and ME expand in the solar wind away from the Sun, the longer duration of the sheath at 1~AU is interpreted as resulting from more solar wind piling up in the sheath from Mercury's to Earth's orbit. Indeed, if the ICME moves away from the Sun with a speed larger than the ambient solar wind, it will continue to accumulate a sheath by a ``snow-plow" effect. Then, the thickness of the sheath depends on both how much plasma and magnetic field is accreted/compressed from the pre-solar wind into the sheath, and how much of it is able to escape toward the sides. This was shown with MHD simulations by \citet{Siscoe2008}: the authors showed that the solar wind piles up in front of ICMEs {{since it is unable to flow completely around the ejecta},} due to the lateral expansion of ICMEs.

Furthermore, physical processes like magnetic reconnection at the rear of an ICME, due to an overtaking stream or another ICME, could also contribute to reducing the size of the ME with respect to an increasing sheath size. The study of \citet{Ruffenach2015} showed that the front and rear boundaries used to define the ejecta are chosen such that the duration of the flux rope is overestimated.
Often, the magnetic ejecta might be reduced in size due to reconnection at the rear, even though it is still expanding overall. Such an erosion effect can however also take place at the front of the ME, therefore leading to the erosion of both the sheath region and the ME. As such, it seems more plausible that the increase of the sheath duration compared with that of the ME is due to a solar wind pile-up as well as the dynamic expansion of the sheath.



\subsection{{Magnetic field correction for the eccentricity of Mercury's orbit}}

ACE and VEX collect data at a roughly constant distance from the Sun, due to the low eccentricity of Earth's and Venus's orbits. On the other hand, Mercury's orbit is more eccentric than that of the other two planets, and as such there is a variation in distance for the sample of data considered. This variation affects the measured ICME magnetic field intensity. 
{It is well established that the magnetic} field intensity within the magnetic ejecta decreases with distance away from the Sun as a power-law \citep{Kumar96,Bothmer1998,Wang2005, Leitner2007, Gulisano2010, Winslow2015}. 
The multi-spacecraft study of \citet{Winslow2015} reports a fit of a power law to the data for the mean ME magnetic field of: 
$<B> = \mathrm{(7.5\pm1.2)}\mathrm{r}^{(-1.95\pm0.19)}$, where $B$ is expressed in nT and $r$ in AU.
The value found in this study for $B$ at 1AU is similar to the ones found by other authors \citep[e.g.,][]{Wang05}, with a steeper radial dependence.
In the following, the magnetic field intensity for all MESSENGER ICME events is normalized so as to consider the same distance from the Sun (taken as 0.37 AU) following the relationship found by \citet{Winslow2015}.\\

%
%

\section{{Superposed epoch results of the entire datasets at the different spacecraft}}
\label{sect_overallSE}

\subsection{Magnetic field intensity profiles at the different spacecraft}

\begin{figure}[t!]    
\centering
\includegraphics[width=0.7\textwidth, clip=]{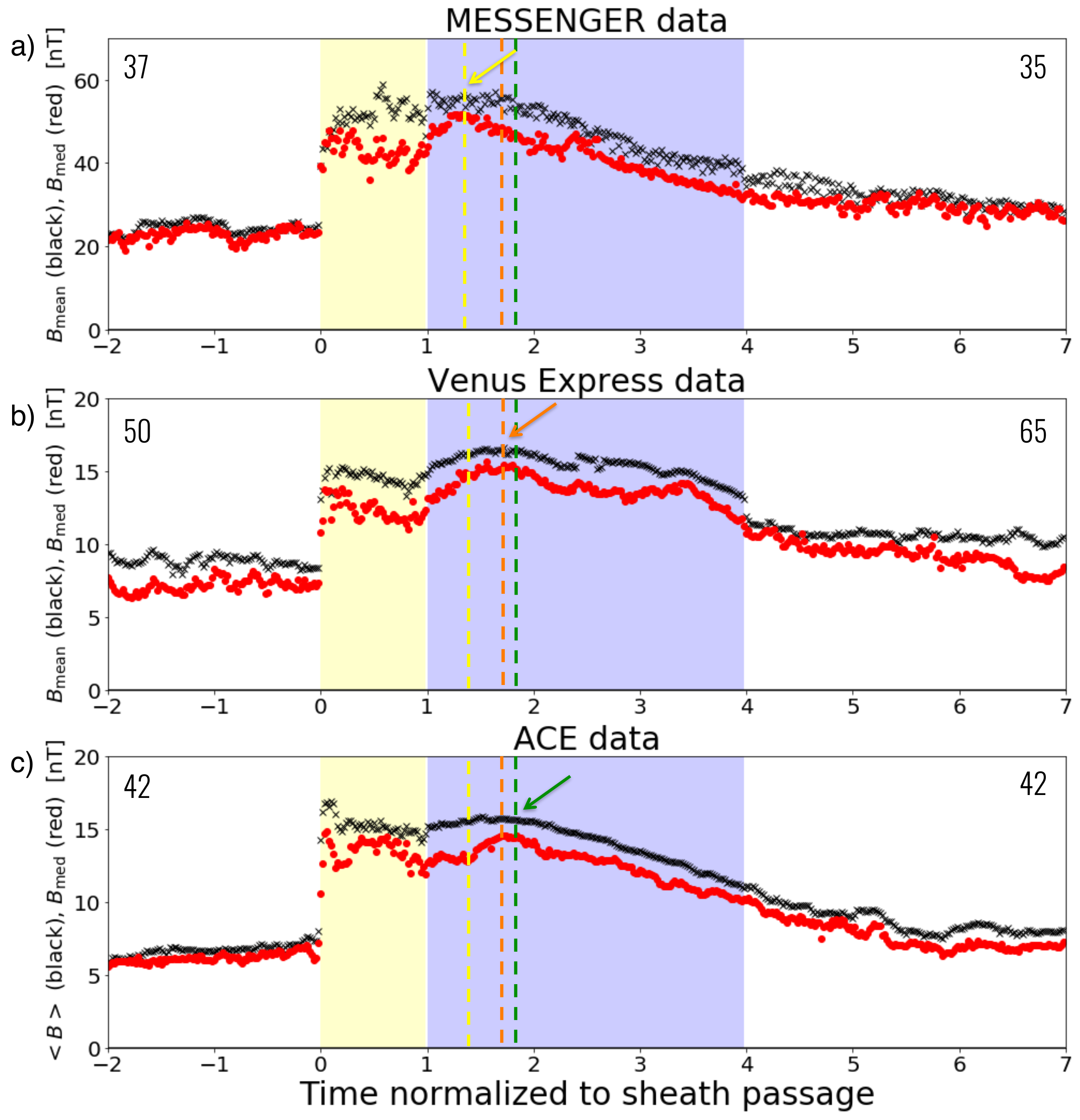}
\caption{Superposed epoch analysis of the mean (black crosses) and median (red points) magnetic field intensity for all the ICMEs seen at MESSENGER (top), Venus Express (middle) and ACE (bottom). The yellow and blue regions are the sheath and the ME regions, respectively (a color convention kept for the superposed epoch plots). The number of studied ICMEs is indicated in the top left for the sheath, and top right for the ME. The normalized time (horizontal axis) is described in \sect{normbin}. The vertical axis provides the field strength in nT. The colored arrows and vertical lines (yellow for MESSENGER, orange for VEX and green for ACE) indicate the locations of the maxima in the median magnetic field profiles. {The relative locations of the vertical lines show an asymmetry that is} more pronounced at MESSENGER that at VEX and ACE.}
\label{fig_SEALL}
\end{figure}

The results obtained for the superposed epoch analyses at different spacecraft are shown in \fig{SEALL} and the data for each superposed epoch have been put online (see all superposed epoch data: \url{https://figshare.com/s/3e0394e629fbed907152}). The top panel shows the superposed epoch analysis of all ICMEs seen at Mercury, the middle one of all ICMEs seen at VEX, and finally the bottom panel of all the selected ICMEs seen at ACE. The number of events taken to obtain the superposed epochs is indicated in each graph. Note that the number of events for the sheath and the magnetic ejecta is not the same. This is because in some cases, even though an ICME is clearly identified, either the sheath or the ME lacks enough data points so that these substructures do not enter into the averaging calculation for the superposed epoch. The timeline has been normalized so that $t_{\mathrm{norm}}=0$ corresponds to the {discontinuity} time while on the vertical axis, we plot the magnetic field $B$ in nT. The yellow colored area corresponds to the sheath region, while the blue one corresponds to the magnetic ejecta region. {The sheath duration is normalized to one time unit, and the magnetic ejecta to three time units. This ratio, representative of the results found in \sect{size_ratio}, is set at all spacecraft so as to provide a more straightforward comparison between the different averaged profiles.

In all these superposed epochs, the transition between the sheath and the ME region at $t_{\mathrm{norm}}=1$ is less clearly seen than the transition at $t_{\mathrm{norm}}=0$, but a change from a decreasing to an increasing total field is evident. The transition between the ME and the wake is somewhat continuous.}

\begin{table}[h!]
\centering
\begin{tabular}{| c || c | c | c |}
\hline
\multicolumn{4}{|c|}{Magnetic field intensities (in nT)} \\
\hline
Substructures &  MESSENGER & VEX & ACE \\
\hline
$\Bsw$ & 23.0 ($22.9 \pm 1.4$) & 7.2 ($7.2 \pm 0.4$) & 6.1 ($6.2 \pm 0.3$) \\ 
$\Bs$ & 42.6 ($42.9 \pm 2.7$) & 12.2 (12.3 $\pm 0.7$) & 13.7 (13.4 $\pm 0.9$)\\
$\Bm$&  42.8 ($42.1 \pm 5.8$) & 13.7 (13.8 $\pm 0.9$) & 12.9 ($12.5 \pm 1.3$) \\
$\Bw$ &  29.8 ($29.8 \pm 1.7$) & 9.4 (9.3 $\pm 0.9$) & 7.2 ($7.7 \pm 1.0$)\\
\hline
\end{tabular}
\caption{Median values, as well as the mean and the standard deviation (in brackets) of the total magnetic field intensities within the solar wind, the sheath, the ME and the wake for each of the superposed epoch obtained at MESSENGER, VEX and ACE}
\label{tableBintensities}
\end{table}

We quantified the median value of the magnetic field intensity within each of the substructures (sheath and ME) as well as the preceding solar wind and wake region for comparison, for each superposed epoch obtained at the three different spacecraft (Table~\ref{tableBintensities}). The values of the magnetic field intensities decrease from MESSENGER all the way to ACE, which is an expected result as discussed in \sect{size_ratio}. 
The discontinuity jump, calculated using the ratio of the intensities between the sheath and the solar wind preceding the ICME, is 1.85 at MESSENGER, 1.70 at VEX and 2.23 at ACE: the ratio therefore stays at comparable values.
We also note that the ratio of the ME magnetic field intensity and that of the sheath is 1.01 at MESSENGER, 1.14 at VEX and 0.94 at ACE, which is not that different from one spacecraft to another. Interestingly, for these events the average sheath and ejecta magnetic field strengths are quite similar.

Finally, the wake does not return to the same level of magnetic field intensity as the preceding solar wind after a time interval comparable to the ME interval (the ratio of magnetic field in the wake over that of the pre-ICME solar wind is 1.29 for MESSENGER, 1.31 for VEX and 1.17 for ACE). This indicates that the passage of an ICME is felt much longer than just within the interval of the ICME disturbance. This effect was pointed out recently in the study by \citet{Temmer2017}, who found that interplanetary space needs approximately 2 to 5 days at 1~AU to recover from the impact of ICMEs.

\subsection{Evolution of the magnetic field asymmetry}

At Mercury's orbit, the ME total magnetic field {increases and decreases shortly after the start of the ME. The profiles obtained at the other two spacecraft show that the start of the decrease is further away from the beginning of the ME than for MESSENGER}.
We note however that none of the superposed epoch profiles are completely smooth. This could be due to the variability of each profile and the relatively small number of similar profiles within the samples. 


We found that for all spacecraft, the profiles of the ME show a peak in $|B|$, indicated with the yellow arrow and dashed lines in the MESSENGER data, orange in the VEX data and green in the ACE data (\fig{SEALL}). {The lines have been superposed on the results for all spacecraft to compare the position of the peaks for each spacecraft}. This peak appears closer to the sheath for the MESSENGER data compared with the others, and moves slightly towards the center of the ME for the VEX and ACE data. We also quantify this change in the symmetry of the profile by calculating the first moment of each of the curves delimited by the ME boundaries:

\begin{equation}
\Delta t = \frac{1}{(t_{\mathrm{rear}}-t_{\mathrm{front}})<B>}\int^{t_{\mathrm{rear}}}_{t_{\mathrm{front}}} (t-\frac{t_{\mathrm{rear}}+t_{\mathrm{front}}}{2}) B(t) dt
\label{asymmetryparameter}
\end{equation}

This first moment, written as $\Delta t$ and expressed in units of time, is defined as the center of mass for 1D distributions. It is a quantitative measure used here to quantify the asymmetry of the profile. This is a standard mathematical expression generally used for calculating the weighted mean position of a distribution of mass, or any quantity. It is a simple and robust measure of the asymmetry of a function. Other parameters such as distortion parameters can also be found in the literature \citep{Nieves2018}.  Considering only the superposed epoch profile of the ME for each spacecraft with a duration normalized to 1 (so that the beginning of the ME is at 0 and its end is at 1), we found that the first moment of the MESSENGER curve profile is situated at a distance 0.46 away from the beginning of the ME (0.04 from the center of the time interval), that for the VEX curve is at 0.48, and that for ACE is at 0.47 (respectively 0.02 and 0.03 from the center of the time interval). {The different locations of the first moments for each profile therefore follow the same trend as for the peaks}. Quantifying the asymmetry, the comparison between all the superposed epochs at all the spacecraft does not give a clear indication of an evolution of an asymmetric total field profile with heliocentric distance.

Note also that the superposed epoch for the VEX data shows a second bump within the ME profile (\fig{SEALL}). Looking back at each individual case, we were not able to find a few specific events that could indicate why this bump is there. To check the stability of this bump, we removed several events and looked at their influence on the overall profile. By repeating this with different criteria (removing only outlier events, or removing events with a profile away from the average one, or removing events with structures within the ME), we confirmed that the bump is still present. {We conclude that this second bump in the magnetic profile is not due to a small number of events, that is, outliers with a specific property.}

\subsection{Magnetic ejecta without sheaths}

\begin{figure}[t!]    
\centering
\includegraphics[width=1\textwidth, clip=]{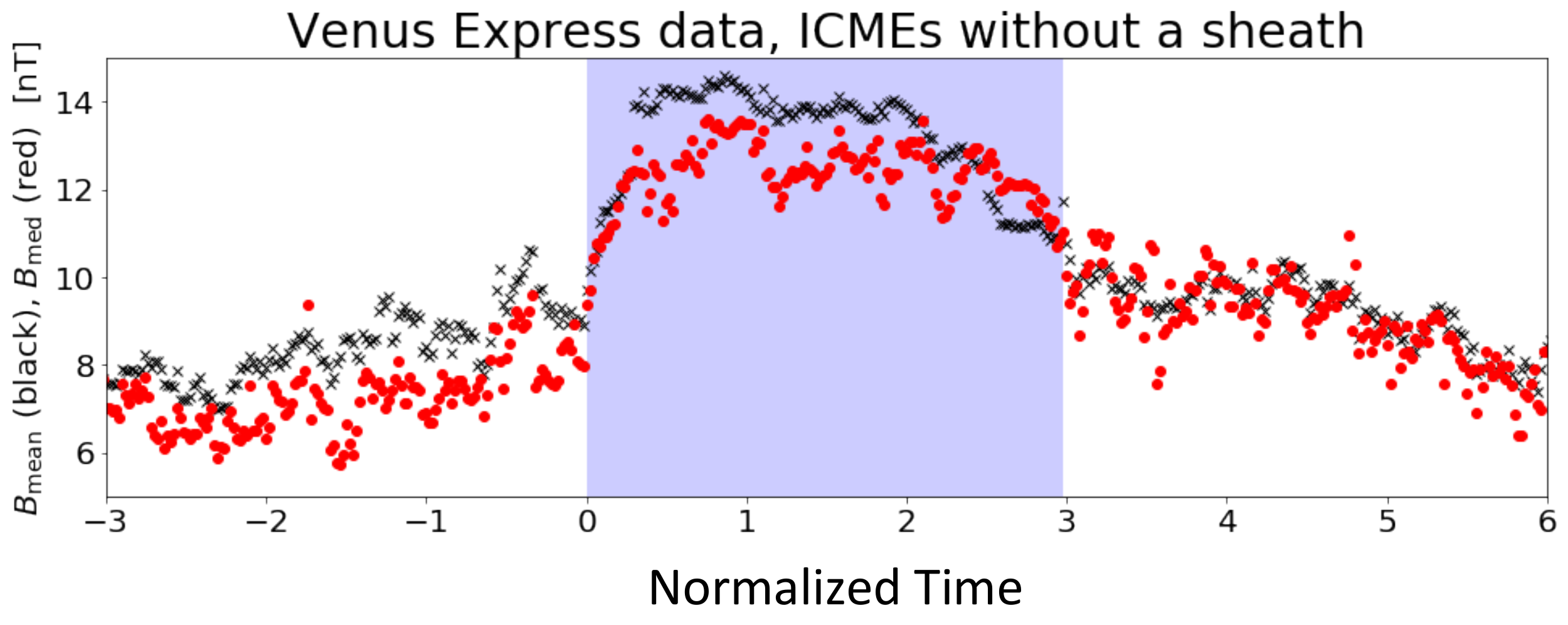}
\caption{Superposed epoch for the 19 ICMEs without a sheath detected by Venus Express. The ME region is indicated with the blue shaded area. The plotting convention is the same as in \fig{SEALL}.}
\label{fig_VEXMEonly}
\end{figure}

We also selected events within the VEX sample that were not associated with any sheath (we found 19 cases). Note that since all ICMEs studied at MESSENGER and ACE have a sheath, we investigated the magnetic ejecta without sheath in the VEX data only.
We computed the superposed epoch for these cases, as shown in \fig{VEXMEonly}. In this figure, the blue shaded area represents the ME region, preceded and followed by the surrounding solar wind. Interestingly, the superposed epoch ME profile has a rounder profile than that of the ME preceded by a sheath region as was shown in \fig{SEALL}b. We calculated the asymmetry (as discussed above) and found that the first moment was located at 0.5, i.e. at the center of the ME. 

Following the results from \citet{Masias2016}, slower ICMEs, which also tend to have weaker sheaths (in terms of the ratio between the solar wind and sheath magnetic field intensity value) display a symmetric profile for the ME substructure. This was interpreted as a consequence of a state in relaxation with the surrounding conditions and little interaction with the sheath. 
The ICMEs without sheaths showing a symmetric profile {at the orbit of Venus extend the results from \citet{Masias2016} obtained at Earth's orbit.} When the sheath is non-existent, the speed at which the structures are propagating is close to that of the solar wind. In such conditions, the magnetic field within the ME only needs to ``adjust" to the surrounding environment without an extra higher total pressure in the sheath, providing this symmetric profile.

%
%

\section{Superposed epoch analyses using subcategories of ICMEs}
\label{sect_subcategories}

In the following, we investigate in more detail the profiles obtained at each spacecraft by separating the samples into subgroups of ICMEs.
\subsection{What can classifying ICMEs reveal?}
In \citet{Masias2016}, the authors analyzed the generic {magnetic field intensity} profiles of ICMEs at 1~AU and the dependence of this profile with the average MC speeds. To do so, they created three subcategories of ICMEs classified with their speeds (low, medium and high speed). They found that the MC inside ICMEs with the lowest speeds have a more symmetric profile, with a less pronounced sheath, while the MCs propagating with a higher speed have a more asymmetric profile with a higher jump of $B$ at the shock preceding the sheath (see their Fig. 4). They concluded that slow MC profiles can be interpreted as resulting from a relaxed force-free configuration, contrary to their faster counterparts. In the following, we investigate whether such differences between different categories of ICMEs can be observed closer to the Sun.

Classifying ICMEs with their average speed is not easy to do for the MESSENGER and the VEX data. First, the parameter is not readily available for both missions and for all cases.  In some cases, it was possible to obtain limited solar wind speed data from MESSENGER's FIPS instrument \citep[e.g. see Fig 2 in][]{Good2015}, however such data are not available for most ICME cases at MESSENGER.  A short burst of density and speed measurements were made in the solar wind by VEX's ASPERA-4 instrument once every 24 hrs \citep[e.g. see Fig. 9 in][]{Rouillard2009}. However, the recovered speed data is very patchy and has considerably larger uncertainties associated with it than data obtained from dedicated solar wind plasma analyzers (as found on ACE, for example).  

Therefore, we investigate in the following the possibility to use another parameter to rank the magnetic clouds in different categories that can correlate well with the speed. One straightforward possibility is to use the transit time from the Sun to estimate the average speed at Mercury and at Venus. The \citet{Winslow2015} list has an estimate of the speed based on the Sun-Mercury propagation, from the identification of the coronal CME causing the ME at MESSENGER. The speeds found for the set of data vary between 300 to 2400~km/s. However, {the associated CME has not been identified for all the events, so the transit speed can only be calculated for some of the events}. \\

\subsection{Sheath magnetic field strength as proxy for the ME speed}
Since the proton speed is available in the ACE 1~AU data, and since we only have the magnetic field intensities available at MESSENGER and VEX, we investigate in the following whether any correlation exists between the speed and the magnetic field intensity recorded within the different substructures.

We start by calculating the Pearson and Spearman correlation coefficients between the mean or the median value of the total magnetic field and the speed inside the sheath and inside the ME for each ICME seen at ACE. We find that it makes no difference whether we use the mean or the median values for this calculation, so we further proceed using the median values.

\begin{figure}[t!]    
\centering
\includegraphics[width=1\textwidth, clip=]{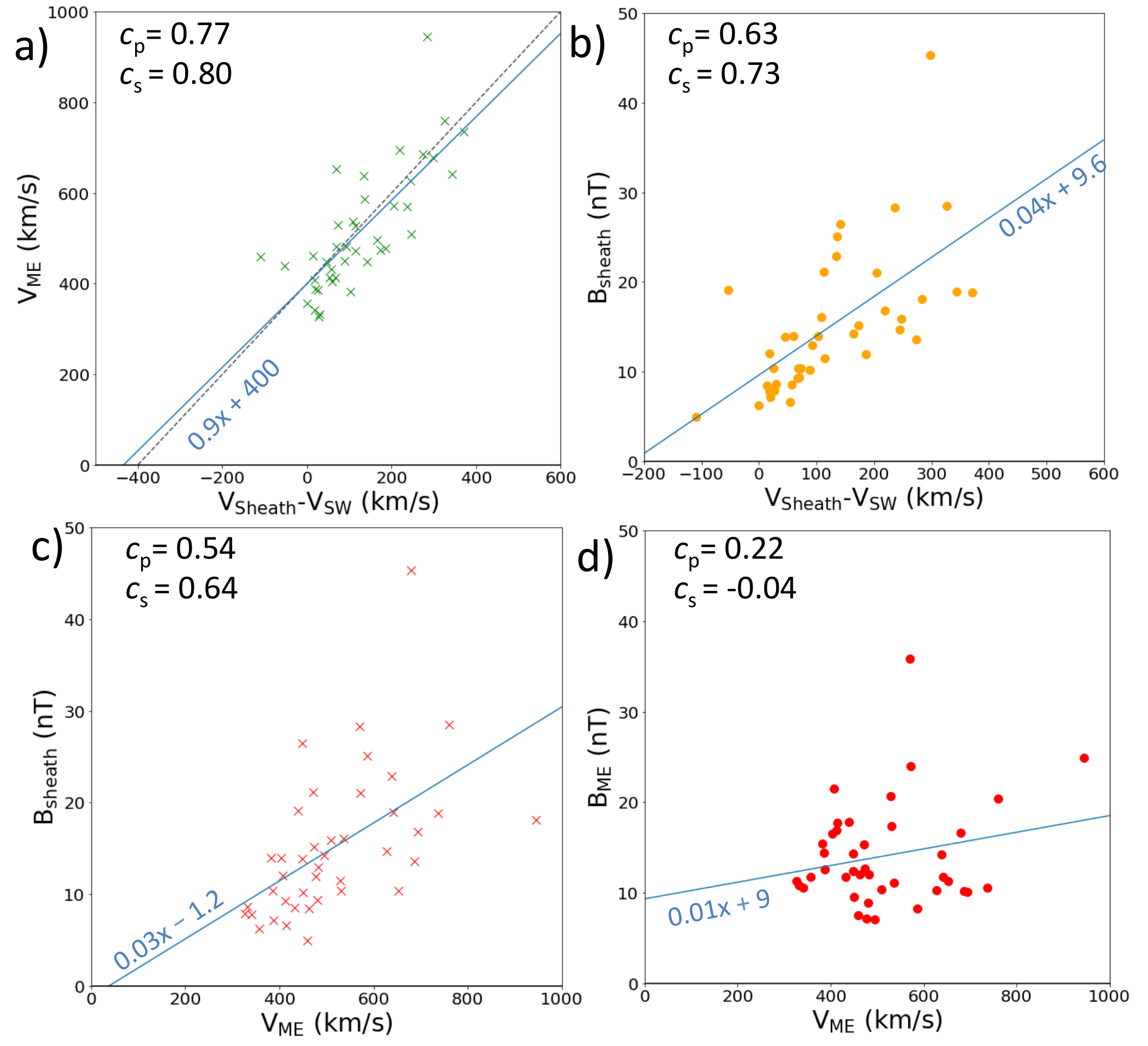}
\caption{Correlations between the speeds and the magnetic fields within the ICME substructures seen at ACE. The Pearson and Spearman correlations are indicated, as well as the linear regression for each graph (in blue). a) Correlation between $\Vm$ and $\Vs-\Vw$. $\Vs$ and  $\Vw$ are the median velocities in the sheath and in the pre-ICME solar wind, where the time interval duration is the same as that for the sheath. The black dashed line represents the identity function. b) Correlation between $\Bs$ and $\Vs-\Vw$. c) Correlation between $\Bs$ and $\Vm$. d) Correlation between $\Bm$ and $\Vm$.}
\label{fig_CorrelationACE}
\end{figure}

The results of four different correlation studies using ACE data are shown in \fig{CorrelationACE} with the Pearson and Spearman correlation coefficients in Table \ref{tablecorrel}. The best correlation is found between the proton speed within the ME ($\Vm$) and within the sheath ($\Vs$, Table \ref{tablecorrel}). This result is expected: as the sheath is being driven by the ME, we indeed expect these substructures to propagate with similar speeds. {Accordingly}, a slightly lower correlation is present between $\Vm$ and $\Vs-\Vw$, {where $\Vw$ is the solar wind speed preceding the ICME} (\fig{CorrelationACE}a). 

We then look at the correlations between the magnetic field intensities within the sheath $\Bs$  ({respectively} within the ME, $\Bm$) with the speeds. The best correlation is found between $\Bs$ and $\Vs-\Vw$ (\fig{CorrelationACE}b). We find that $\Bs$ and $\Vm$ are moderately correlated {having Pearson and Spearman correlation coefficients of 0.54 and 0.64, respectively (\fig{CorrelationACE}c). $\Bm$ and $\Vm$ show poor or even no correlation with Pearson and Spearman correlation coefficients of 0.22 and -0.04, respectively (\fig{CorrelationACE}d).} These correlations are in agreement with the results found in \citet{Owens2005} for a larger sample of ICMEs at 1~AU{{, as well as the results of \citet{Liu2008} who showed that the flow speed and the magnetic field in the sheath correlates well with the speed of the ejecta.}}

Note here that we do not investigate the relation between the maximum magnetic field intensity in the ME and the ICME speeds. These were found to be moderately correlated: \citet{Richardson2010} found that, at least for ICMEs with a magnetic cloud, the correlation coefficient is around 0.6 (see their Fig. 11) while \citet{Mostl2014} later confirmed a relationship between the two (see their Fig. 14). In our study, we mostly focus on correlations with median/mean quantities. This is because maximum values are not representative of the overall magnetic field behavior. Furthermore, maximum values for the magnetic field are not always found for VEX and MESSENGER data due to the magnetosphere crossings. We therefore choose a quantity that can be consistently studied with the different samples available to us, and moreover which is statistically robust (as it is derived from an ensemble of points in contrast to a single point).

{The correlation between $\Bs$ and $\Vm$, means that{, assuming this relationship also holds for ICMEs at Mercury and Venus,} we now have a proxy for categorizing ICMEs with their speeds.} In the following, we therefore define three categories of MESSENGER, VEX and ACE ICMEs with weakest, medium and strongest $\Bs$ values, which we expect to be comparable to classifying the events by increasing order of speed.

\begin{table}[h!]
\centering
\begin{tabular}{| c || c | c |}
\hline
\multicolumn{3}{|c|}{Correlation coefficients} \\
\hline
Parameters &  Pearson & Spearman \\
\hline
$\Bs$ vs $\Bm$ & 0.37 & 0.26 \\ 
$\Vs$ vs $\Vm$ & \textbf{0.92} & \textbf{0.93} \\
$\Vs-\Vw$ vs $\Vm$ & \textbf{0.77} & \textbf{0.80} \\
$\Bm$ vs $\Vm$ & 0.22 & -0.04 \\
$\Bs$  vs $\Vm$ & \textbf{0.54} & \textbf{0.64} \\
$\Bm$ vs $\Vs$  &  0.14 & -0.10 \\
$\Bs$ vs $\Vs$  &  0.46 & \textbf{0.60} \\
$\Bm$ vs $\Vs-\Vw$ & 0.18 & -0.02 \\
 $\Bs$ vs $\Vs-\Vw$ & \textbf{0.63} & \textbf{0.73} \\ 
\hline
\end{tabular}
\caption{Pearson and Spearman correlation coefficients {comparing the median total magnetic field $B$ within each substructures,   $B$ with the median proton speed $V$, as well as $V$ for each substructures}. The ICMEs are sampled at ACE with a 1s time cadence. The highest correlations are indicated in bold.}
\label{tablecorrel}
\end{table}

\subsection{Superposed epoch profiles for sub categories of ICMEs}
\begin{figure}[t!]    
\centering
\includegraphics[width=0.95\textwidth, clip=]{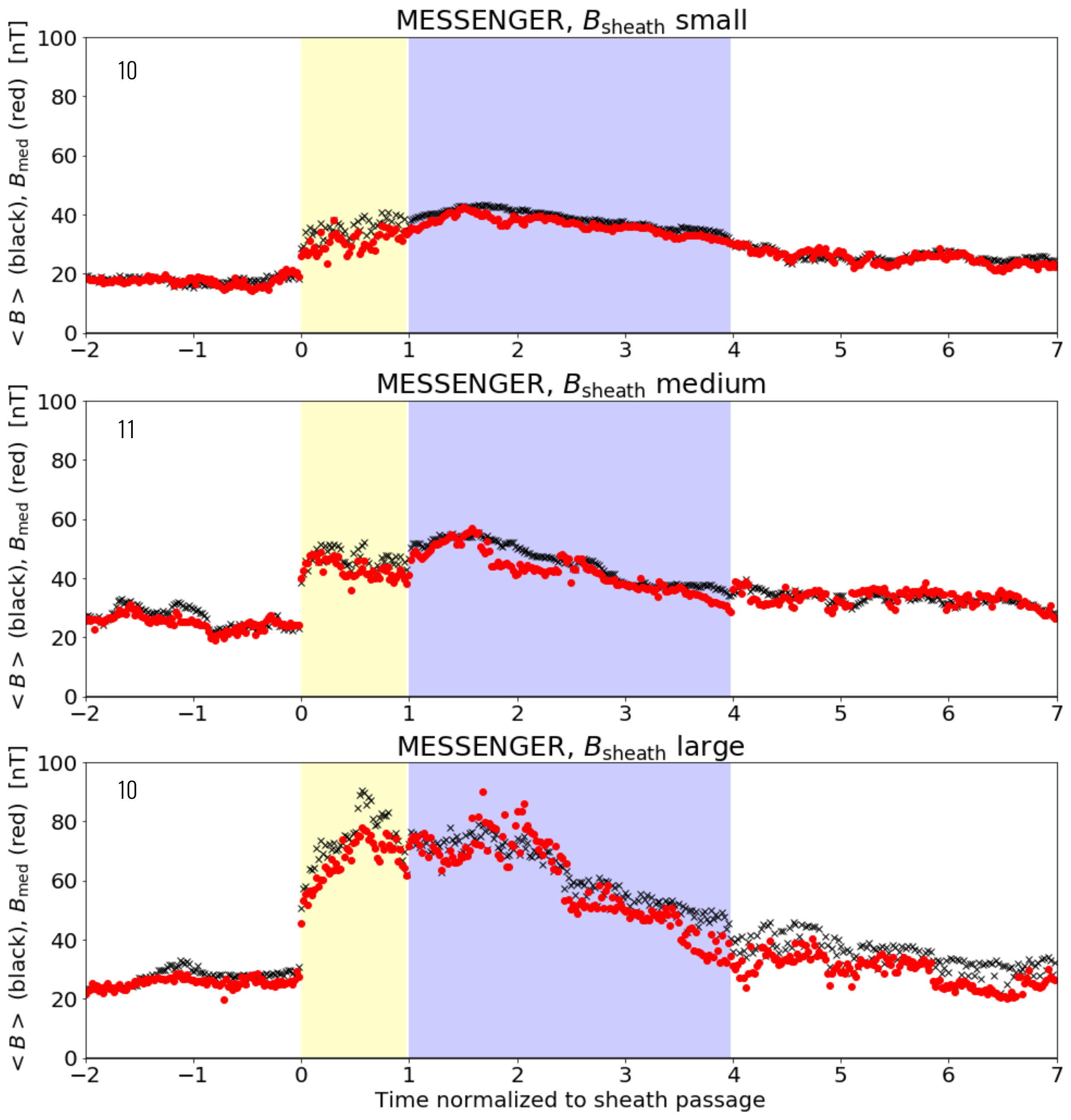}
\caption{Superposed epoch of the total magnetic field for the MESSENGER data with three categories, where all the events are classified with increasing $\Bs$ intensities. The yellow and blue regions are the sheath and the ME regions, respectively (a color convention kept for the superposed epoch plots). The black crosses (resp. red points) are the mean (resp. median) values over the number of studied ICMEs (indicated in the top left). The normalized time (horizontal axis) is described in \sect{overallSE}. The vertical axis provides the field strength in nT. }
\label{fig_SE-MES-cat}
\end{figure}

\begin{figure}[t!]    
\centering
\includegraphics[width=0.95\textwidth, clip=]{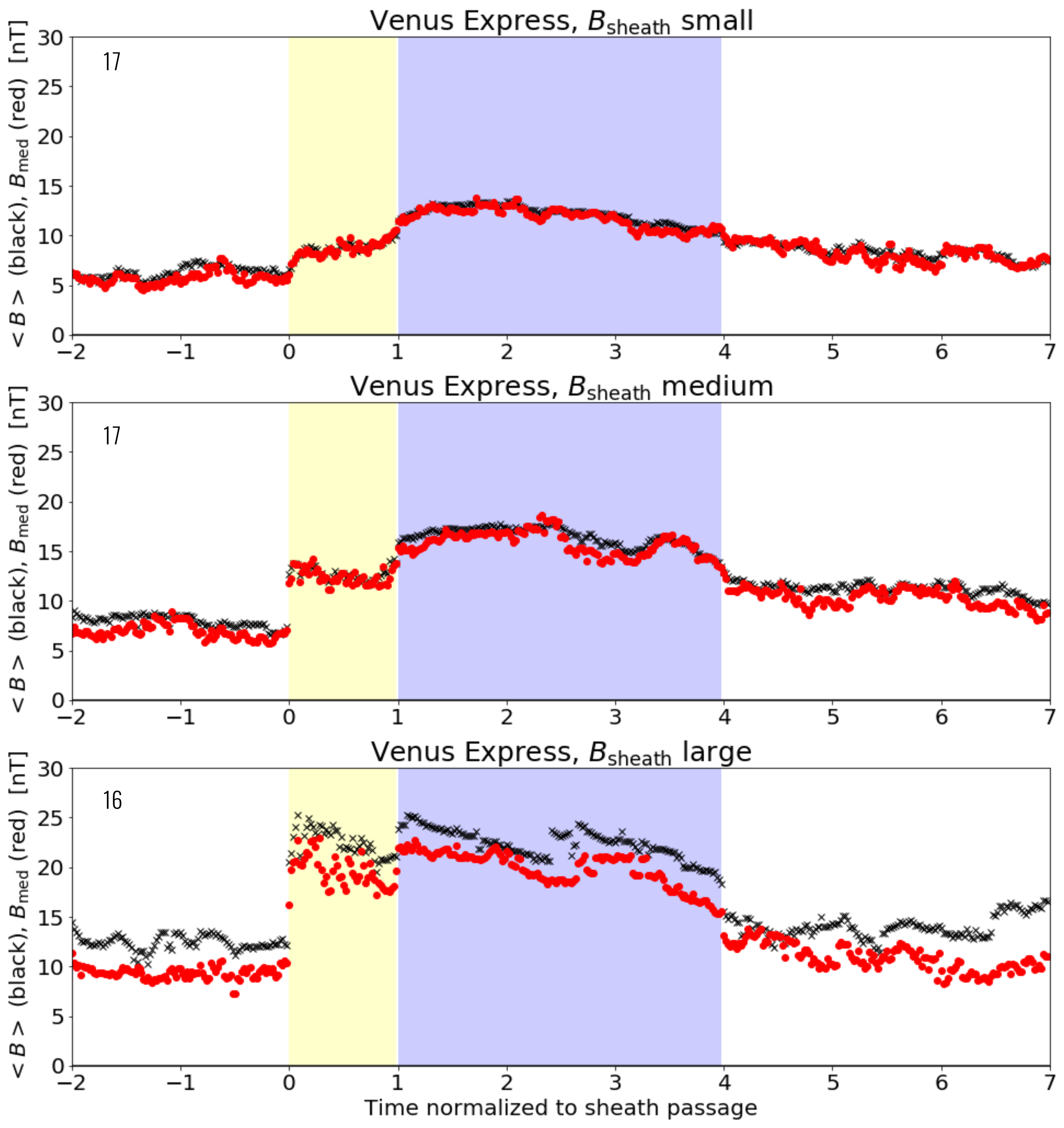}
\caption{Superposed epoch of the total magnetic field for the Venus Express data with three categories, where all the events are classified with increasing $\Bs$ intensities. Same color convention as \fig{SE-MES-cat}.}
\label{fig_SE-VEX-cat}
\end{figure}

\begin{figure}[t!]    
\centering
\includegraphics[width=0.95\textwidth, clip=]{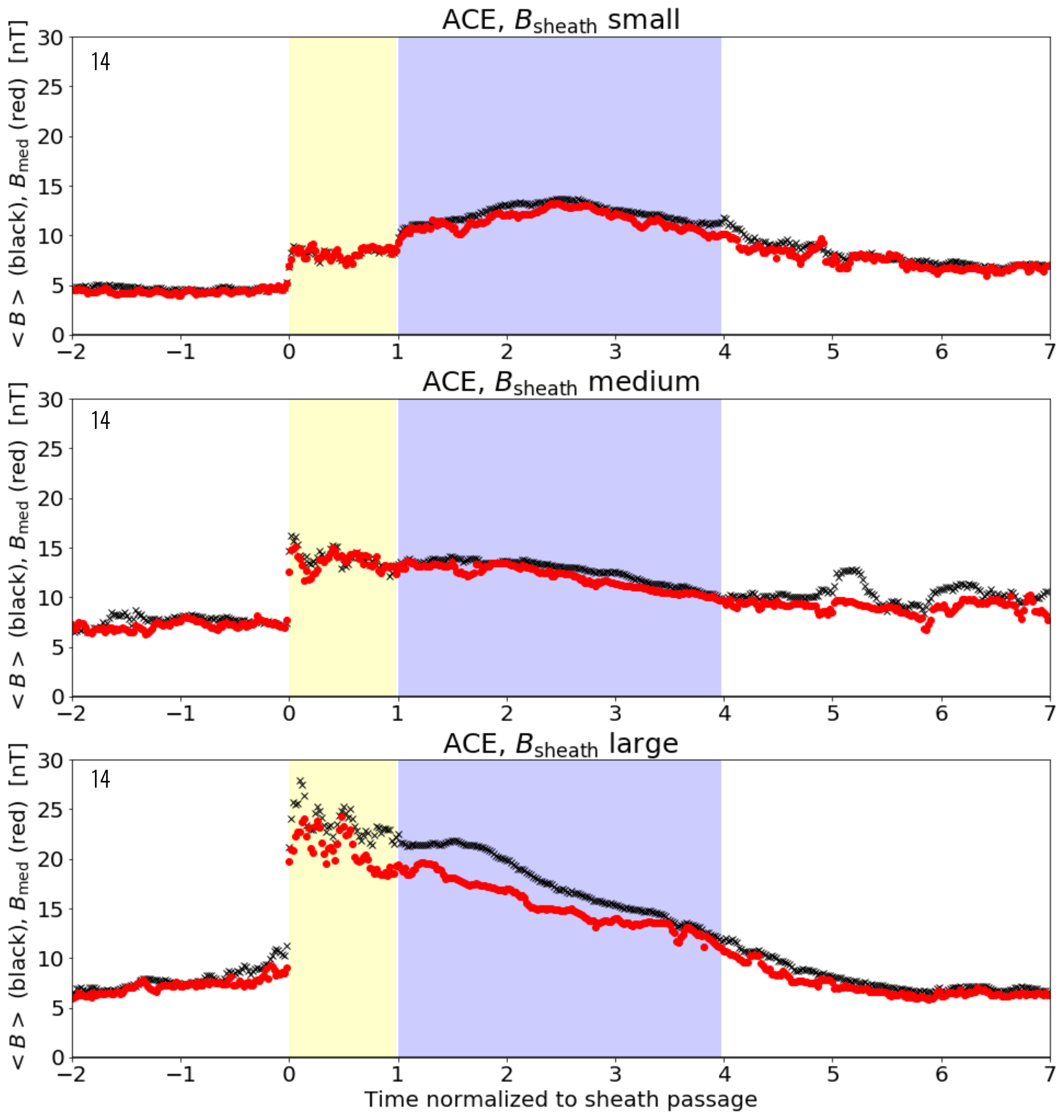}
\caption{Superposed epoch of the total magnetic field for the ACE data with three categories, where all the events are classified with increasing $\Bs$ intensities. Same color convention as \fig{SE-MES-cat}.}
\label{fig_SE-ACE-cat}
\end{figure}

The results of the superposed epoch analysis made for each of the categories, and at each spacecraft, is shown in \fig{SE-MES-cat} for the MESSENGER data, \fig{SE-VEX-cat} for the VEX data and \fig{SE-ACE-cat} for the ACE data.

\subsubsection{Profiles for each category}
Overall, the same trend can be found between the different spacecraft: the superposed epoch profile with ICMEs with lower $\Bs$ has a flatter profile {(while still peaked)} at all spacecraft, while the profiles from larger $\Bs$ show a steeper decline in the magnetic field strength from the leading to the trailing edge of the ME. Note that while we used $\Bs$ as a proxy to classify the ICMEs, we find that the three subcategories for ACE data show the same profiles as the ones found in \cite{Masias2016} when the classification was made on the speed within the ME, therefore justifying the use of  $\Bs$ as an ordering parameter. 


For the small $\Bs$ category, we find that the profile of the ME is more asymmetric at Mercury's orbit than it is at Earth's orbit, with a peak of the profile close to the sheath for MESSENGER data while the peak is clearly at the center for ACE data, so an evolution towards an equilibrium. We provide below a quantification of this asymmetry.

\subsubsection{Magnetic field intensities}
\label{sect_magneticfieldintensities}
{As a next step, we compare} the magnetic field intensities in all the substructures of the ICME superposed epoch and at all spacecraft. The results for the solar wind magnetic field intensities are reported in the first line in Table \ref{tableratiocategories}. We find a similar trend between the same categories at all spacecraft for the characteristics {of the solar wind preceding the ICME: the magnetic field strength of the solar wind ($\Bsw$) preceding ICMEs having large $\Bs$ (i.e.\ faster ICMEs) is higher than the magnetic field strength of the solar wind preceding ICMEs having small $\Bs$ (i.e.\ slower ICMEs).} This is not simply a correlation between $\Bsw$ and $\Bs$ since the same trend was found with categories defined with $\Vm$ in \citet{Masias2016}.



We next checked the ratio of the magnetic field intensities between the different substructures for each of the categories. The results are given in the second to fourth line in Table~\ref{tableratiocategories}. 

We find first that the ratio of the sheath magnetic field intensity and that of the pre-solar wind, $\Bs / \Bsw$, increases between the category ``small'' (with the lowest $\Bs$) and category ``large'' (with the highest $\Bs$), from 1.7 to 2.5 for the MESSENGER data, from 1.4 to 2 for the VEX data, from 2 to 2.5 for the ACE data{, as expected from the category selection}.


We also find that the {ratio $\Bm / \Bs$ decreases from category ``small'' to category ``large'' for each of the spacecraft (from 1.1 to 0.8 for MESSENGER data, from 1.4 to 1.1 for VEX data, and from 1.4 to 0.7 for the ACE data).} For both MESSENGER and ACE, we also find 
that the magnetic field intensity in the sheath region is larger than that inside the ME for the ICMEs within category ``large'' (\fig{SE-MES-cat}, \fig{SE-ACE-cat}). This stronger sheath is consistent with the asymmetric magnetic field profile in the ME.

Finally, we also calculate the ratio of magnetic field intensities between {the wake and the pre-ICME solar wind regions. We find that at all spacecraft, the ratio decreases (from 1.4 to 1.1 for the MESSENGER data, from 1.4 to 1.1 as well for the VEX data, and from 1.7 to 0.9 for the ACE data)}. In other words, the solar wind magnetic field intensities pre- and post- ICMEs are more different for ``slower'' ICMEs than for ``faster'' ones. While ICMEs detected as fast at different spacecraft are ICMEs that are originally fast and that keep a higher speed relative to the ambient medium, slower ICMEs are ICMEs that are {in a range from originally slow to} fast events that have been slowed down by the solar wind{, so as to become all slow ICMEs when observed in situ} \citep[see][]{Masias2016}. As such, the differences in the pre- and post- ICME solar wind characteristics for slow events could be due to {initially faster} ICMEs that have drastically been slowed down. For such cases, interactions with the solar wind (such as drag) could be accompanied with magnetic reconnection eroding the magnetic flux rope of the ejecta, leading to a larger wake dominated by this eroding mechanism and as such with different characteristics than the pre-ICME solar wind. Also, slow ICMEs are more likely to be overcome by a fast solar wind stream or a fast ICME. In such cases, the wake region would be disturbed by these structures, which furthermore could be compressed by the interaction \citep[such an effect was shown in][]{Rodriguez2016}.

\begin{table}[h!]
\centering
\begin{tabular}{| c || c | c | c || c | c | c || c | c | c |}
\hline
& \multicolumn{3}{c ||}{  MESSENGER } & \multicolumn{3}{c || }{  VEX } & \multicolumn{3}{c| }{  ACE } \\
\hline
$\Bs$ & small & medium & large  & small & medium & large  & small & medium & large  \\
\hline
$\Bsw {(nT)} $ &  17.7 & 25.2 & 25.6 & 5.7 & 6.8 & 9.4 & 4.3 & 7.2 & 7.3\\
$\Bs / \Bsw$ &  1.7 & 1.7 & 2.5 & 1.4 & 1.7 & 2 & 2 & 2 & 2.5\\
$\Bm / \Bs$ &  1.1 &  1 & 0.8 & 1.4 &  1.2 & 1.1 & 1.4 & 0.9 & 0.7  \\
$\Bw / \Bsw$ &  1.4 & 1.4 & 1.1 & 1.4 & 1.4 & 1.1 & 1.7 & 1.2 & 0.9 \\
Asymmetry parameter & -0.02 & -0.04 & -0.07 & -0.02 & -0.01 & -0.02 & 0 & -0.03 & -0.05\\ 
\hline
\end{tabular}
\caption{{Values of the magnetic field intensity in the solar wind and} ratio of median magnetic field intensities in each substructure for each subcategories of ICMEs seen at different spacecraft and classified with the median magnitude of $\Bs$. }
\label{tableratiocategories}
\end{table}

\subsubsection{Asymmetry parameter}
We calculated the asymmetry parameter in the same way as \sect{overallSE} (see Eq. \ref{asymmetryparameter}), the results being shown in the last line of Table \ref{tableratiocategories}. We find a similar tendency between the different categories for the MESSENGER and ACE data: the asymmetry parameter {is negative as expected with a stronger field at the front. Its absolute value, which indicates the significance of the asymmetry, increases} from 0.02 for the ``slower'' ICMEs to 0.07 for the ``faster'' ICMEs for MESSENGER, and from 0 (i.e. a central peak) to 0.05 for the ACE data. {The superposed epoch profiles for VEX are less smooth, having more pronounced bumps than for MESSENGER and ACE. They also} imply a different conclusion: we found approximately the same asymmetry parameter for all the ICME categories. However, when looking at \fig{SE-VEX-cat}, one can see that the profiles for each category are qualitatively similar to that found for MESSENGER and ACE {if we do not take into account the second (later) peak for ``medium'' and ``faster'' categories.}

In summary, we find that the ``slower'' ICMEs have a more symmetric shape, with a peak closer toward the middle of the profile, while for ``faster'' ICMEs, the profile steepens, with a peak closer to the sheath. 

We conclude that the slower MEs relax with increasing solar distance,  which is {related to the existence of} a weaker sheath and a longer time to reach a given solar distance. In contrast, the fast MEs {{and their expansion}} drive a strong shock, then a strong sheath pressure builds up, which creates asymmetric conditions between the front and the rear of faster MEs. These events being still significantly faster than the overtaken solar wind, the sheath is expected to keep being built up to at least 1 AU. 
Also, slower ICMEs are more likely to be overtaken by fast streams or another CME, which creates a compression at the rear of the structure. This could affect the profile to make it more symmetric by creating an increase of the magnetic field intensity at the rear of the ME similar to that found at the beginning of the ME (fourth row of Table \ref{tableratiocategories}). {Understanding such mechanisms will require further case studies and associated modelling.}

\subsection{Does aging impact the profiles of ICMEs?}

Comparing MESSENGER and ACE data means that we are comparing ICMEs at different ``ages'', with the age of the structure corresponding to the time it has spent in the solar wind. {Since the ICME takes a finite time to pass the observing spacecraft, the leading edge of the ICME is observed in a younger stage than the trailing edge.} 


One can estimate an average age difference between when the ICME starts (at the front of the ICME) and ends (at the rear of an ICME). The sampling at MESSENGER starts approximately 20 hours after the front of the ICME has left the coronal environment of the Sun. This has been estimated using the transit speed of all ICMEs seen departing the Sun in our sample and detected at MESSENGER near Mercury.
The typical magnetic ejecta time duration in our MESSENGER sample is around 10 hours. This means that the observation of the rear of the ICME is made 30 hours after the structure has left the solar corona, so when the ICME is 1.5 times older than the front. 

Similarly, when a typical ICME arrives at ACE, its front part has left the Sun's atmosphere approximately 3.5 days before. Since the typical ICME duration is estimated around 30 hours in our sample, this means that the rear part of the ICME is around 1.36 times older than its front, so similar to what was found for MESSENGER. 

In \citet{Demoulin2008} and \citet{Demoulin09c}, the authors investigated models of ICME expansion {and its consequences} to understand the profiles of ICMEs.  A spacecraft measures the magnetic field at different times across the encountered ICME.  If the ICME is evolving during the crossing, for example expanding, the observed data are mixing the spatial and temporal variations.  
The presence of an expansion implies a transformation of the observed field profile, because the younger front of the ICME is more compact than the older rear. This is detected as a higher magnetic field intensity at the front than at the back, as we found for the ``fast'' ICME category at all spacecraft. This effect is larger for larger ICMEs as they are observed during a longer time (for the same mean velocity value). 
However, \citet{Demoulin2008} showed that this effect was not significant enough to explain the asymmetry found in the ICME profiles at 1~AU. 

We found, in our limited sample, that the aging is similar at Mercury to what it is at Earth: ICMEs take around 3 times more time to propagate from the Sun to Earth than to Mercury, but also take 3 times longer to be sampled at Earth than at Mercury, in agreement with a comparable non-dimensional expansion rate found in the inner-heliosphere (with the HELIOS missions,  \citealp{Gulisano2010}) and at 1~AU. Then, we conclude that the aging effect likely does not have an important effect on the asymmetry profile seen at Mercury. {More precisely, at Mercury as at 1~AU, the aging effect is expected to create a significant asymmetry in $B$ between the front and the rear of the ME only within very large ICMEs expanding faster than usual.}



\subsection{Are superposed epoch analyses robust to the definition of ICME boundaries?}
\label{boundaries_analysis}

Defining the boundaries of ICMEs is difficult (see \sect{size_ratio}). On the one hand, it is highly dependent on the availability of different sets of data (e.g. plasma and magnetic field parameters) that can help to consider wider criteria and guide the eye when defining the different substructures of ICMEs. On the other hand, defining boundaries can also depend from one person to another, as these boundaries are not objectively and clearly defined. To check the robustness of the superposed epoch analysis, we checked whether the results were consistent from one catalog of boundaries to another.

As described in \sect{datasets}, we first used the catalogs already available for the MESSENGER, VEX and ACE ICMEs and we had a second look at all of the events listed. While the disturbance start is relatively easy to define, as the transition between the pre-ICME solar wind and the sheath is clear, the transition between the sheath and the magnetic ejecta, as well as the transition between the magnetic ejecta the post-ICME solar wind, are both more difficult to define. In this second check, we reassessed the boundaries by eye and then compared our results. We then constructed a list of ICMEs with the boundaries defining the shortest magnetic ejecta found between the different co-authors of this paper, as well as a list with the less restrictive boundaries. The reason for choosing the shortest durations for the magnetic ejecta is to be the most conservative in terms of defining the magnetic ejecta.
When applying the superposed epoch technique, we found very little difference between the superposed epochs built from the two different lists, therefore showing that the method is robust and is not impacted by the fluctuations linked with the subjective definition of boundaries. This means that the results of the superposed epoch analysis were not strongly dependent on the exact choice of the boundaries.
The list presented in Appendix A is the most conservative (most restrictive) list of ICME boundaries.

All events in the catalogs can also be categorized depending on the quality of the data, most specifically on how easy it is to define an ICME. Indeed, some events cannot be fully sampled by MESSENGER or VEX because of magnetospheric crossings. Also, since the superposed epoch technique averages different events, it is important to check whether outliers in our sample (e.g. time series corresponding to higher-than-average magnetic field intensities) can affect the results. Although not shown here, we did a careful analysis removing different events to check the robustness of the results. Since on average we only found up to 4 outlier events, the superposed epoch profiles, especially those with the variance,  were not affected by the removal of these outliers. As such, the technique is robust to a few outliers and in previous sections, the results are presented with the whole dataset (Appendix A) for each spacecraft sample.

%
%

\section{Conclusion}
\label{sect_conclusion}

This paper presents a comparison between statistical studies performed using magnetic field data from different spacecraft, namely MESSENGER, VEX and ACE. These three spacecraft sampled ICMEs at different distances from the Sun, from the orbit of Mercury ($\approx 0.4$), to that of Venus (0.72~AU) and Earth (1~AU).  The conclusions are listed below.

\begin{itemize}
\item{We first investigated the typical duration of the magnetic ejecta and the sheath. We found that this duration increases with solar distance for both substructures, due to their expansion when the ICME moves away from the Sun, while the ratio of the duration of the magnetic ejecta over that of the sheath decreases from MESSENGER to ACE.
We propose that, as ICMEs continue their propagation away from the Sun, the sheath size increases due to the accretion of overtaken plasma and magnetic field. This could be investigated more quantitatively with mission data that provide in situ measurements of the speed.}
\item{We applied the superposed epoch method to obtain generic ICME magnetic field intensity profiles. 
which provides insights on their typical features, as they are enhanced by the averaging method. 
The method was applied to all events in each sample (\sect{overallSE}).
We quantified the change in the asymmetry of profiles and found a slightly higher asymmetry at Mercury compared to the other two heliospheric locations (Venus and Earth).
}
\item{The superposed epoch analysis shows that at any distance from the Sun, the magnetic field intensity in the wake (the solar wind region following the ICME within the same time scale of the ME) does not fully recover the properties of the pre-ICME solar wind. This is in agreement with the finding by \citet{Temmer2017} that the solar wind has a long recovery period after the passage of an ICME.}
\item{By separating ICMEs without a sheath in the Venus Express data, we found that the magnetic ejecta had a symmetric profile. This is coherent with the conclusion {made at 1 AU} of \citet{Masias2016} who showed that slow ICMEs tend to have a weaker sheath and a more symmetric profile.}
\end{itemize}

Due to the strong variability in the ICME profiles at each spacecraft, we decided to investigate how to create subcategories of ICMEs. Because the speed is only partially available for MESSENGER and VEX, we used $\Bs$, the median magnetic field intensity in the sheath, which appears to be a good proxy of the magnetic ejecta speed based on ACE results. We assume that this result also applies to the MESSENGER and VEX ICMEs, {and these results are backed up by the ACE data for which the in situ proton velocity is available \citep{Masias2016}}. The results of the superposed epoch for each categories of ICMEs at each spacecraft are listed below:

\begin{itemize}
\item{The solar wind magnetic field intensity level in front of ICMEs is weaker (resp. higher) for weak (resp. strong) $\Bs$ ICMEs. This could be due to the solar origin of the ICMEs, where ICMEs with strong $\Bs$ originate from regions of the Sun where the magnetic fields of both the ICME and the pre-ICME are strong.}
\item{The magnetic fields in the sheath region and in the ME are higher for ``fast'' ICMEs compared with ``slow'' ICMEs. This could be due to the shock compression, {{as well as the dynamic pressure due to the magnetic ejecta behind which are both higher in fast events \citep[e.g.][]{Manchester2005}}}. This then increases the magnitude of the magnetic field, which evolution is not fully transmitted to the ME (both substructures being out of equilibrium).}
\item{The solar wind {magnetic field intensities post-ICMEs are larger than the pre-ICME ones especially for the slower ICMEs.  This difference between slow and fast ICMEs} may be linked with {a more frequent and stronger} interaction with an overtaking solar wind for slow ICMEs.}
\item{The ME profiles are more symmetric for ``slow'' ICMEs (i.e. with a weaker $\Bs$), while those classified as ``faster'' (i.e. with a stronger $\Bs$) have a sloped profile and a larger asymmetry, with a larger magnitude of the magnetic field at the front than at the rear of the ME. }
\end{itemize}
The category of ICMEs classified as ``faster'' with the $\Bs$ proxy {are probed when they are younger} than their ``slower'' counterparts. Having spent less time in the solar wind, they had less time to adjust the conditions within the ICMEs with the surrounding solar wind. In other words, these structures are less relaxed and therefore are expected to display a less symmetric magnetic field profile, which is what we have found for each category of ``fast'' ICMEs at the three spacecraft. {On the other hand, the interaction with the sheath is stronger for the faster events, which also contributes to the asymmetry.}

The present study provides insights on the evolution of the profiles of selected similar subsets of ICMEs at different heliospheric distances. The mechanisms behind the different profiles we have found, especially with regard to the speed of ICME propagation, will need to be investigated. Especially, MHD codes simulating the ICME propagation within the inner heliosphere will be invaluable in understanding the conditions needed to recover the results we have found with in situ data. Furthermore, future missions such as ESA's Solar Orbiter and NASA's Parker Solar Probe will be essential in completing catalogs of in situ detected ICMEs and in providing new datasets necessary to investigate the evolution of ICMEs in the interplanetary medium.

\appendix

\section{Tables of ICMEs seen at MESSENGER, Venus Express and ACE}
\label{sect_AppendixA}

\begin{sidewaystable}
\scriptsize
\centering
\begin{tabular}{l l l l l l l l l l}
Event & Year of Event & Discontinuity\footnote{either a disturbance or a shock}  & Discontinuity & ME\footnote{``ME'' stands for magnetic ejecta} leading Edge & ME leading Edge & ME trailing edge & ME trailing edge & $r_H$\footnote{Heliospheric distance of the s/c in AU}  & Catalog    \\    
number &  & DOY & Time & DOY & Time  & DOY & Time & AU & reference\footnote{``G'' stands for \citep{Good2015}, ``W'' stands for \citep{Winslow2015}}    \\    \hline
1   & 2009 & 19   & 22:30:00 & 20   & 06:04:19 & 20  & 23:54:14 & 0.444 & G    \\    
2   & 2009 & 45   & 22:00:00 & 46   & 04:23:31 & 46  & 14:12:28 & 0.327 & G    \\    
3   & 2009 & 51   & 16:15:00 & 51   & 19:36:28 & 52  & 00:20:09 & 0.371 & G    \\    
4   & 2009 & 266 & 11:55:00 & 266 & 13:55:12 & 267 & 06:53:16 & 0.343  & G   \\  
5   & 2009 & 360 & 16:33:36 & 360 & 16:33:36 & 361 & 18:21:36 & 0.439 &  G    \\  
6   & 2010 & 242 & 16:53:45 & 242 & 22:00:00 & 243 & 12:37:26 & 0.378  & G    \\  
7   & 2010 & 309 & 11:45:36 & 309 & 18:00:00 & 310 & 13:07:40 & 0.462  & G   \\  
8   & 2010 & 342 & 22:55:00 & 343 & 02:55:00 & 343 & 15:40:00 & 0.337  & G    \\  
9   & 2010 & 347 & 04:16:19 & 347 & 11:44:09 & 347 & 16:45:07 & 0.365 &  G    \\  
10 & 2011 & 68   & 02:08:09 &  68  & 02:08:09 & 68   & 12:28:48 & 0.331 &  G   \\  
11 & 2011 & 139 & 11:50:02 &  139 & 16:45:07 & 140 & 02:00:00 & 0.413  & G, W   \\
12 & 2011 & 155 & 15:12:04 & 155 & 16:15:46 & 156 & 00:43:24 & 0.325 &  W    \\    
13 & 2011 & 156 & 03:31:27 & 156 & 04:29:45 & 156 & 06:10:00 & 0.323  & G, W    \\    
14 & 2011 & 159 & 04:25:27 & 159 & 06:45:00 & 159 & 10:06:41 & 0.313 &  W    \\    
15 & 2011 & 172 & 18:01:48 & 172 & 21:30:00 & 173 & 14:55:00 & 0.333  & W    \\    
16 & 2011 & 265 & 10:24:32 & 265 & 12:46:24 & 265 & 21:41:37 & 0.358 &  W \\
17 & 2011 & 278 & 02:32:45 & 278 & 06:02:02 & 278 & 12:32:19 & 0.427 &  W    \\
18 & 2011 & 288 & 08:26:36 & 288 & 13:30:00 & 289 & 06:23:02, & 0.460  & G, W     \\  
19 & 2011 & 364 & 16:27:23 & 364 & 21:12:57 & 365 & 09:19:52 & 0.420  & G, W     \\    
20 & 2012 & 2     & 18:28:13 & 2    & 19:54:10 & 2      & 22:50:00 & 0.434 &  W    \\   
21 & 2012 & 2     & 22:55:00 & 3    & 00:42:00 & 3      & 04:53:00 & 0.434 &  new   \\    
22 & 2012 & 3     & 04:52:48 & 3    & 07:53:57 & 3      & 12:40:37 & 0.436  & W    \\    
23 & 2012 & 37   & 23:30:03 & 38  & 00:49:59 & 38    & 09:08:51 & 0.415  & W    \\   
24 & 2012 & 64   & 10:34:41 & 64  & 14:36:30 & 64    & 19:00:00 & 0.309  & W    \\    
25 & 2012 & 64   & 20:00:00 & 64  & 22:30:00 & 65    & 12:30:00 & 0.309 &  new   \\   
26 & 2012 & 65   & 12:28:50 & 65  & 13:15:39 & 65    & 15:15:00 & 0.311  & W    \\    
27 & 2012 & 67   & 04:37:44 & 67  & 06:10:53 & 67    & 14:52:41, & 0.315 &  G, W     \\   
28 & 2012 & 71   & 05:34:10 & 71  & 06:31:14 & 71    & 10:00:00 & 0.331  & W    \\   
29 & 2012 & 264 & 18:29:00 & 264 & 19:17:00 & 264 & 21:18:00 & 0.426  & W    \\    
30 & 2013 & 248 & 13:39:59 & 248 & 15:15:35 & 248 & 22:54:41 & 0.416  & W    \\   
31 & 2013 & 299 & 11:07:22 & 299 & 13:21:07 & 299 & 23:15:00 & 0.349  & W   \\    
32 & 2013 & 300 & 12:35:00 & 300 & 15:00:00 & 301 & 01:53:06 & 0.344 &  W    \\    
33 & 2013 & 302 & 11:14:46 & 302 & 11:14:46 & 302 & 19:10:55 & 0.334  & W   \\   
34 & 2014 & 24   & 03:25:19 & 24   & 04:25:10 & 24  & 08:15:00 & 0.340  & W   \\    
35 & 2014 & 27   & 18:37:50 & 27   & 19:55:25 & 28  & 00:00:00 & 0.323  & W    \\    
36 & 2014 & 208 & 16:01:06 & 208 & 19:10:14 & 209 & 06:06:00 & 0.309  & W    \\    
37 & 2014 & 245 & 08:00:31 & 245 & 08:27:50 & 245 & 12:38:52 & 0.454  & W    \\    
38 & 2014 & 348 & 08:55:29 & 348 & 09:52:29 & 348 & 17:19:15 & 0.463  & W    \\  
39 & 2015 & 80   & 09:42:02 & 80   & 11:28:46 & 80   & 15:03:10 & 0.439 &  W    \\   
40 & 2015 & 97   & 23:30:30 & 98   & 02:01:56 & 98   & 06:16:39 & 0.351  & W    \\   
41 & 2015 & 104 & 16:05:39 & 104 & 17:41:27 & 104 & 22:27:30 & 0.318 & W    \\    \hline
\end{tabular}
\caption{Table of ICMEs seen at MESSENGER}
\end{sidewaystable}

\begin{sidewaystable}
\scriptsize
\centering
\begin{tabular}{l l l l l l l l l}
Event & Year of Event & Discontinuity$^{4}$ &  Discontinuity & ME$^{5}$ leading Edge & ME leading Edge & ME trailing edge & ME trailing edge &  $r_H^6$    \\    
number &   & DOY & Time & DOY &  Time & DOY & Time & AU     \\    \hline
1   &  2007 & 44 & 04:48:00 & 44 & 14:40:16 & 45 & 09:33:13 & 0.725   \\    
2   &  2007 & 117 & 00:14:24 & 117 & 08:52:48 & 117 & 16:10:33 & 0.719  \\   
3   &  2007 & 126 & 00:43:12 & 126 & 08:24:00 & 126 & 20:52:48 &  0.719 \\   
4   &  2007 & 144 & 19:12:00 & 145 & 03:36:00 & 145 & 19:59:31 &  0.721 \\   
5   &  2007 & 167 & 02:15:21 & 167 & 02:15:21 & 167 & 17:16:48 & 0.724  \\    
6   &  2007 & 167 & 22:19:12 & 167 & 23:31:12 & 168 & 10:42:14 & 0.724  \\   
7   &  2007 & 285 & 17:24:00 & 285 & 18:43:12 & 285 & 21:34:33 &  0.722  \\      
8   &  2007 & 321 & 04:19:12 & 321 & 07:20:38 & 321 & 19:00:28 & 0.719  \\       
9   &  2007 & 341 & 18:08:38 & 342 & 03:44:38 & 343 & 14:44:09 &  0.719  \\   
10 &  2008 & 364 & 16:13:26 & 364 & 20:45:36 & 365 & 04:58:04 &  0.723  \\   
11 &  2009 & 17 & 13:33:36 & 17 & 14:36:14 & 17 & 22:49:43 &  0.721  \\     
12 &  2009 & 118 & 21:38:52 & 119 & 14:24:00 & 120 & 05:08:09 & 0.725  \\    
13 &  2009 & 136 & 10:19:12 & 136 & 21:37:26 & 137 & 16:22:04 &  0.727   \\    
14 &  2009 & 153 & 16:04:48 & 153 & 18:38:52 & 154 & 12:20:09 &  0.728  \\    
15 &  2009 & 158 & 05:58:33 & 158 & 05:58:33 & 158 & 14:34:04 &  0.728  \\    
16 &  2009 & 175 & 04:33:36 & 175 & 05:45:36 & 175 & 11:45:36 &  0.728    \\    
17 &  2009 & 191 & 07:59:31 & 191 & 10:37:55 & 192 & 06:24:28 &  0.727    \\    
18 &  2009 & 221 & 09:15:50 & 221 & 10:27:50 & 221 & 19:16:19 &  0.723    \\    
19 &  2009 & 290 & 05:45:36 & 290 & 05:45:36 & 290 & 23:06:43 &  0.719   \\    
20 &  2009 & 325 & 09:50:24 & 325 & 09:50:24 & 325 & 15:59:02 &  0.722     \\    
21 &  2010 & 62 & 04:37:03 & 62 & 15:39:27 & 63 & 03:35:51 &  0.725     \\    
22*&  2010 & 74 & 15:36:00 & 74 & 15:36:00 & 75 & 12:28:48 &  0.724  \\
23 &  2010 & 75 & 18:57:36 & 76 & 09:18:34 & 77 & 12:14:15 &  0.724    \\    
24 &  2010 & 157 & 13:39:21 & 157 & 14:35:31 & 158 & 00:44:38 &  0.719   \\    
25 &  2010 & 166 & 14:09:36 & 166 & 23:21:07 & 168 & 01:43:40 & 0.72     \\    
26 &  2010 & 173 & 14:52:48 & 173 & 14:52:48 & 174 & 18:28:48 & 0.721     \\    
27 &  2010 & 180 & 00:57:36 & 180 & 08:38:24 & 180 & 23:00:57 & 0.722   \\    
28 &  2010 & 213 & 14:41:16 & 213 & 23:11:02 & 214 & 05:55:40 &  0.726    \\    
29 &  2010 & 214 & 11:29:45 & 214 & 12:41:45 & 214 & 16:04:48 &  0.726     \\    
30 &  2010 & 214 & 19:42:14 & 214 & 22:06:14 & 215 & 14:29:45 & 0.726    \\    
31 &  2010 & 222 & 03:48:57 & 222 & 12:31:40 & 222 & 21:14:24 & 0.727    \\    
32 &  2010 & 224 & 22:19:12 & 224 & 22:19:12 & 225 & 05:19:40 &0.727    \\    
33 &  2010 & 251 & 11:03:50 & 251 & 11:03:50 & 251 & 18:33:07 &  0.728    \\    
34 &  2010 & 253 & 16:07:40 & 253 & 21:57:36 & 254 & 17:08:09 & 0.728    \\    
35 &  2011 & 81 & 08:51:21 & 81 & 17:28:01 & 82 & 18:20:44 &  0.727     \\    
36 &  2011 & 101 & 10:26:24 & 101 & 11:03:50 & 101 & 17:32:38 & 0.728    \\    
37 &  2011 & 107 & 07:39:21 & 107 & 07:39:21 & 107 & 18:17:16 &  0.728     \\    
38 &  2011 & 111 & 11:19:40 & 111 & 11:19:40 & 112 & 11:26:52 &  0.728   \\    
39 &  2011 & 139 & 20:58:33 & 140 & 03:51:50 & 140 & 18:12:57 & 0.727   \\    
40 &  2011 & 156 & 05:16:48 & 156 & 08:38:24 & 156 & 22:30:43 &  0.724    \\    
41 &  2011 & 182 & 11:31:12 & 182 & 13:45:07 & 183 & 09:01:26 &  0.721     \\    
42 &  2011 & 273 & 23:16:48 & 274 & 03:36:00 & 274 & 14:42:43 &  0.723   \\    
43 &  2011 & 289 & 00:50:24 & 289 & 06:05:45 & 290 & 09:38:52 &  0.725    \\    
44 &  2011 & 359 & 12:38:52 & 359 & 15:38:52 & 360 & 00:46:04 &  0.727    \\    
45 &  2011 & 360 & 23:16:48 & 361 & 02:19:40 & 361 & 15:41:45 &  0.727    \\    
46 &  2012 & 32 & 16:48:00 & 32 & 21:07:37 & 33 & 13:24:48 &  0.722    \\    
47 &  2012 & 61 & 13:37:55 & 61 & 21:26:38 & 62 & 13:21:04 &  0.719 \\
48 &  2012 & 67 & 13:26:24 & 67 & 20:13:12 & 68 & 11:42:43 & 0.719 \\
49 &  2012 & 123 & 00:38:52 & 123 & 15:12:57 & 125 & 01:32:09 &  0.722 \\
50 &  2012 & 200 & 16:37:55 & 200 & 23:51:21 & 201 & 16:43:40 & 0.728 \\
51 &  2012 & 211 & 09:50:24 & 211 & 10:59:31 & 211 & 17:58:33 & 0.728 \\
52 &  2012 & 257 & 04:43:40 & 257 & 11:44:09 & 258 & 03:17:16 & 0.722 \\
53 &  2012 & 315 & 15:38:52 & 315 & 21:56:09 & 316 & 01:55:12 &  0.719 \\
54 &  2012 & 318 & 10:48:00 & 318 & 17:31:12 & 319 & 06:27:21 &  0.719 \\
55 &  2012 & 330 & 01:01:55 & 330 & 05:16:48 & 330 & 11:19:40 &  0.72 \\
56 &  2013 & 8 & 09:23:02 & 8 & 15:24:32 & 9 & 19:47:32 & 0.725 \\
57 &  2013 & 33 & 02:15:56 & 33 & 02:15:56 & 33 & 20:51:38 & 0.728 \\
58 &  2013 & 48 & 10:55:20 & 48 & 10:55:20 & 48 & 20:50:47 &  0.728 \\
59 &  2013 & 52 & 16:03:21 & 52 & 16:03:21 & 52 & 19:55:12 & 0.728 \\
60 &  2013 & 65 & 13:22:04 & 65 & 23:35:48 & 66 & 11:38:06 &  0.728 \\
61 &  2013 & 72 & 20:41:16 & 72 & 20:41:16 & 73 & 07:15:36 &  0.727 \\
62 &  2013 & 117 & 14:38:24 & 118 & 01:45:07 & 119 & 13:53:45 &  0.722 \\
63 &  2013 & 201 & 10:06:14 & 201 & 21:10:04 & 202 & 17:47:02 &  0.721 \\
64 &  2013 & 261 & 11:41:16 & 261 & 11:41:16 & 261 & 22:32:09 & 0.728 \\
65 &  2013 & 278 & 04:17:45 & 278 & 12:33:07 & 279 & 14:58:33 & 0.728 \\
66 &  2013 & 334 & 04:30:43 & 334 & 13:59:31 & 335 & 16:00:28 & 0.723 \\
67 &  2013 & 348 & 03:07:12 & 348 & 14:32:38 & 348 & 23:39:50 &  0.721 \\    \hline
\end{tabular}
\caption{Table of ICMEs seen at Venus Express (* added to the list of Good et al)}
\end{sidewaystable}

\begin{sidewaystable}
\scriptsize
\centering

\begin{tabular}{l l l l l l l l l}
Event & Year of Event & Discontinuity$^{4}$ &  Discontinuity & ME$^{5}$ leading Edge & ME leading Edge & ME trailing edge & ME trailing edge    \\    
number &   & DOY & Time & DOY &  Time & DOY & Time   \\    \hline
1 & 1998 & 	63 & 10:55:12	 & 63 & 14:19:25	&	65 & 05:00:57   \\    
2 &  1998	 & 121 & 21:25:55 & 	122 & 11:31:12	&	123 & 16:30:43  \\    
3  & 1998	 & 267 & 23:13:55 & 	268 & 09:25:55	&	269 & 12:25:55  \\    
4  & 1998 & 	291 & 19:01:55 & 	292 & 04:07:12	&	292 & 13:07:40   \\   
5  & 1998	 & 312 & 04:20:38 & 		313 & 00:25:55	&	314 & 00:25:55  \\
6  & 1999	 & 49 & 02:05:15  & 		49  & 	16:17:45	& 50 & 11:16:48  \\
7 & 1999 & 106 & 10:39:21 & 106 & 19:14:51 & 107 & 20:15:21 \\						
8  & 1999	 & 220 & 17:39:50 & 		221 & 19:58:04	&	222 & 15:57:36  \\
9 &  2000	 & 42 & 23:11:01 & 		43 & 16:17:45	&	43 & 23:16:48  \\
10 &  2000 & 	51 & 20:38:24	 & 	52 & 08:55:39	&	53 & 10:56:37  \\
11  & 2000 & 	175 & 12:21:36 & 		176 & 07:16:47	&	177 & 19:16:19  \\
12  & 2000 & 	210 & 05:42:43 & 		210 & 20:06:43	&	211 & 09:07:12  \\
13  & 2000 & 	224 & 18:04:19 & 		225 & 06:16:18	&	226 & 02:16:19  \\
14  & 2000 & 	261 & 16:56:38 & 		262 & 01:01:55	&	262 & 17:02:24  \\
15  & 2000 & 277 & 00:08:38	 & 	277 & 16:14:52	&	278 & 13:14:51  \\
16 &  2000 & 	302 & 09:02:51 & 		302 & 20:09:36	&	303 & 21:08:38  \\
17 &  2000 & 	311 & 09:12:57 & 		311 & 22:09:27	&	312 & 17:24:00  \\
18 &  2001 & 	94 & 14:13:55 & 		94 & 18:01:40	&	95 & 07:20:37  \\
19 &  2001 & 	101 & 13:21:21 & 		102 & 07:30:43	&	102 & 17:29:45  \\
20 &  2001 & 	111 & 15:30:14 & 		111 & 23:30:43	&	113 & 00:30:14  \\
21 &  2001 & 	118 & 04:30:43 & 		119 & 01:30:43	&	119 & 12:30:14  \\
22 &  2001 & 	147 & 14:13:55 & 		148 & 11:13:55	&	149 & 09:12:57  \\
23 &  2001 & 	304 & 12:47:31 & 		304 & 19:59:01	&	306 & 09:00:00  \\
24 &  2002 & 	107 & 10:20:38 & 		108 & 02:15:21	&	109 & 01:14:52  \\
25 &  2002 & 	109 & 08:03:50 & 		110 & 11:28:19	&	111 & 14:28:19  \\
26 &  2002 & 	143 & 10:09:07 & 		143 & 22:20:38	&	144 & 16:20:38  \\
27 &  2002 & 	213 & 04:19:12	 & 	213 & 10:49:26	&	213 & 21:48:57  \\
28 &  2002 & 	213 & 22:17:45 & 		214 & 08:06:43	&	214 & 20:06:43  \\
29 &  2003 & 	79 & 04:14:51	 & 	79 & 11:32:37 &		79 & 21:31:39  \\
30 &  2003 & 	229 & 13:40:48 & 		230 & 10:16:19		& 231 & 04:16:19  \\
31 &  2003 & 	324 & 07:27:50 & 		324 & 10:51:36		& 325 & 01:20:38  \\
32 &  2004 & 	94 & 09:00:00 & 		95 & 00:59:02	&	96 & 13:59:31  \\
33 &  2004 & 	204 & 09:50:24 & 		204 & 14:13:55	&	204 & 20:13:55  \\
34 &  2004 & 	208 & 22:23:31 & 		209 & 01:35:02	&	209 & 11:35:31  \\
35 &  2004 & 	242 & 09:30:14 & 		242 & 19:17:45	&	243 & 19:41:44  \\
36 &  2004 & 	312 & 17:55:39 & 		313 & 04:14:24	&	313 & 16:26:24  \\
37 &  2005 & 	140 & 01:59:31 & 		140 & 06:00:00 &	141 & 03:59:02  \\
38 &  2005 & 	163 & 06:44:38 & 		163 & 16:14:37	&	164 & 05:55:40  \\
39 &  2005 & 	165 & 17:49:55 & 		166 & 04:13:26	&	167 & 08:13:55  \\
40  & 2005 & 	198 & 00:33:07 & 		198 & 12:57:36	&	199 & .1229977  \\
41 &  2005 & 	364 & 23:19:39 & 		365 & 13:53:43 &	1 & 10:19:12  \\
42  & 2006 & 	348 & 13:48:00 & 		348 & 21:36:00	&349 & 19:36:27 \\    \hline
\end{tabular}
\caption{Table of ICMEs seen at ACE}
\end{sidewaystable}

\acknowledgments
The authors thank both referees for their comments which improved the readability of the manuscript. We thank the MESSENGER, Venus Express and ACE mission teams for providing the data that were necessary for the present study. The authors also thank the HELCATS consortium for providing catalogs of ICMEs seen by MESSENGER and Venus Express, as well as I. Richardson and H. Cane for providing a list of ICMEs at 1~AU.  R.M.W. acknowledges support from NASA grant NNX15AW31G and NSF grant AGS1622352. Supports by the Austrian Science Fund (FWF): P31265-N27 and P26174-N27 are acknowledged by T. A. and C.M. S.D. acknowledges support from UBACyT (UBA), PICT-2013-1462, and PIDDEF 2014/8. P.D. thanks the Programme National Soleil-Terre of the CNRS/INSU for financial support.

{The data used in the present paper can be found without any restriction as follows: the MESSENGER magnetic field data are available on the Planetary Data System (\url{https://pds.nasa.gov/}), the Venus Express magnetic field data are available in the ESA's Planetary Science Archive (at \url{https://www.cosmos.esa.int/web/psa/venus-express}). The data from ACE are available at \url{http://www.srl.caltech.edu/ACE/ASC/level2/}. The ICME catalogs used in the study come from different published researcher papers: The MESSENGER ICMEs come from \citet{Winslow2015},  \citet{Winslow2017} and \citet{Good2016}. The Venus Express ICME catalog is provided by \citet{Good2016}. The ACE ICME catalog used here come from \citet{Masias2016}. All superposed epoch data from this paper have been made available online at \url{https://figshare.com/s/3e0394e629fbed907152}}.


%

%






\end{document}